\documentclass{achemso}
\usepackage[english]{babel}
\usepackage{subcaption}
\usepackage[dvipsnames]{xcolor}
\usepackage{natbib}
\usepackage{graphicx}
\usepackage{amsmath}
\usepackage{amssymb}
\usepackage[normalem]{ulem}
\usepackage[version=4]{mhchem}
\usepackage{csquotes}
\usepackage{enumitem}
\usepackage{hyperref}

\usepackage{siunitx} 
\DeclareSIUnit\angstrom{\protect \text{Å}}

\title{Quantifying surfactant adsorption at fluid interfaces by combining X-ray reflectivity and simulations}

\author{Kay-Robert Dormann}
\email{kay-robert.dormann@pkm.tu-darmstadt.de}
\affiliation{Institute for Condensed Matter Physics, Technische Universität Darmstadt, Germany}
\author{Joshua Reed}
\affiliation{Institute for Condensed Matter Physics, Technische Universität Darmstadt, Germany}
\author{Chen Shen}
\affiliation{Deutsches Elektronen-Synchrotron DESY, Hamburg, Germany}
\author{Daniel Mitlewski}
\affiliation{Institute for Condensed Matter Physics, Technische Universität Darmstadt, Germany}
\author{Matej Kandu\v{c}}
\affiliation{Jo\v{z}ef Stefan Institute, Ljubljana, Slovenia}
\author{Benno Liebchen}
\affiliation{Institute for Condensed Matter Physics, Technische Universität Darmstadt, Germany}
\author{Emanuel Schneck}
\email{emanuel.schneck@pkm.tu-darmstadt.de}
\affiliation{Institute for Condensed Matter Physics, Technische Universität Darmstadt, Germany}
\begin{document}
\maketitle

\section{Abstract}
Adsorption of surfactants to fluid interfaces occurs in numerous daily-life and technological contexts. The surfactant surface coverage $\Gamma$ governs interface characteristics like tension $\gamma$, viscoelastic properties, and the stability of thin foam films. Directly measuring $\Gamma$ as a function of the bulk concentration $c$ is highly desirable but challenging, particularly for non-ionic surfactants that lack easily detectable labels. Neutron reflectometry is currently the only generally applicable method, but it is not available for routine experiments. Here, we propose a simulation-assisted approach to deduce the adsorption isotherm $\Gamma(c)$ from X-ray reflectivity data: As a first step, we use atomistic molecular dynamics simulations of surfactant-loaded air/water interfaces with prespecified $\Gamma$ to obtain interfacial electron density profiles. From these profiles, we compute theoretical X-ray reflectivity curves and compare them with experimental measurements to determine the matching bulk concentration. We focus on two non-ionic surfactants (\ce{C12EO6} and $\beta$-\ce{C12G2}) with previously established force fields to illustrate how this combined approach of experiments and simulations can determine the adsorption isotherm. Additional insights are gained through comparison with the measured surface tension isotherms $\gamma(c)$, based on the equation of state $\gamma(\Gamma)$ from simulations.\\

\textbf{Keywords:} monolayer, surface tension, critical micelle concentration, adsorption models

\section{Introduction}
Surfactants are small amphiphilic molecules that accumulate at air/water and oil/water interfaces. They thereby lower the interfacial tension and are commonly used as cleaning agents and to stabilize foams and emulsions~\cite{stubenrauch2004stability, georgieva2009link, langevin2023recent}. Because surfactants are technologically so important~\cite{myers2020surfactant}, with annual production in the billion-dollar/euro range, great efforts are being made for the rational design of surfactants with particular characteristics and functionalities. On the fundamental level, various thermodynamic models have been developed over the past decades~\cite{frumkin1925electrocapillary, liggieri1999molecular, fainerman2019particular} to describe the experimentally most accessible quantities, like the interfacial tension $\gamma$ and the interfacial viscoelastic moduli, both as functions of the surfactant bulk concentration $c$. In all these models, the surfactant excess $\Gamma$ at the interface, i.e., the concentration-dependent number of adsorbed surfactants per surface area, is a central parameter. Unfortunately, the \emph{adsorption isotherm} $\Gamma(c)$ cannot routinely be quantified experimentally. What is typically measured instead is the \emph{surface tension isotherm} $\gamma(c)$. In thermal equilibrium, the two concentration-dependent quantities $\gamma$ and $\Gamma$ are coupled through the Gibbs equation,
\begin{equation}
\Gamma(c)=-\frac{1}{k_{\text{B}}T}\frac{\text{d}\gamma(c)}{\text{d}\ln{(c)}},
\label{eq:gibbs}
\end{equation}
where $k_{\text{B}}T$ is the thermal energy. However, as noted earlier~\cite{li1999neutron, kovalchuk2023surfactant, schneck2024experimental}, this relation is often of limited practical use and can involve ambiguities, such that an independent experimental determination of $\Gamma$ is required. For example, small systematic errors in the tensiometric measurements at low $c$ can affect the apparent slope in $\gamma(c)$. Limitations also arise in other situations, such as the formation of sub-layers, ionizable surfactants with $\Gamma$-dependent protonation degree, under non-equilibrium steady-state conditions~\cite{manning1998measurement, campbell2004external}, or when assumptions have to be made regarding the coupling between ionic surfactants and their counterions~\cite{butt2023physics}.\\
The most direct way to determine the surfactant excess involves the reflection of neutrons, X-rays, and light~\cite{schneck2024experimental}. Neutron reflectometry (NR) can be considered the gold standard~\cite{staples1993surface, lu1994neutron, li1999neutron, xu2013limitations, li2013application, li2014limitations} and is most sensitive when deuterated surfactants are used in combination with so-called non-reflecting water~\cite{staples1993surface, lu1994neutron, li1999neutron}. NR has revealed, for instance, that the predictive power of the Gibbs equation with regard to $\Gamma(c)$ strongly varies between different surfactant types~\cite{xu2013limitations, li2013application, li2014limitations}. Such measurements are relatively fast, and the required deuterated versions of surfactants have become increasingly available. Nevertheless, the limited beamtime on the few horizontal reflectometers at powerful neutron sources worldwide is essentially reserved for cutting-edge soft-matter research of fundamental character. The technique will therefore not be available for routine work on surfactants in the foreseeable future. Ellipsometry with visible light is fast and simple but relies on the approximation of a homogeneous and isotropic surfactant refractive index and requires independent calibration, e.g., from independent NR measurements~\cite{manning1998measurement, valkovska2003measurement}. For surfactants with covalently-bound unique elements like phosphorus or sulfur or ionic surfactants with quasi-bound counterions, total-reflection X-ray fluorescence (TRXF)~\cite{bloch1985concentration, yun1990x} offers an elegant route to determining $\Gamma$~\cite{daillant1991interaction, schneck2024experimental}. However, this technique is inapplicable to non-ionic surfactants without unique chemical elements, which constitute a substantial fraction of the surfactants nowadays utilized. X-ray reflectometry (XRR)~\cite{Als-Nielsen2011, daillant2008x} and the related grazing-incidence X-ray scattering (GIXS) technique~\cite{mora2004x, wiegart2005thermodynamic, shen2024reconstructing} are powerful tools to characterize fluid interfaces structurally in terms of their electron density profile. So far, they have, however, almost exclusively been applied to water-insoluble surfactants~\cite{als1989x, rieu1995melting}, to lipids~\cite{kuhl1999packing}, or with additional information from NR~\cite{Sloutskin2022}. $\Gamma$ is in these cases typically known \emph{a~priori} from surface diffraction in case of crystalline arrangements~\cite{hermelink2008unsaturated} or from well-calibrated pressure/area isotherms. For water-soluble surfactants, which usually do not assume ordered structures and for which pressure/area isotherms cannot be obtained, $\Gamma$ can only be deduced from the interfacial electron density profiles themselves. But this is non-trivial and usually ambiguous because electron-density contributions from surfactants and displaced water molecules must be accurately disentangled. For this reason, the only conclusive XRR studies have dealt with (per-)fluorinated surfactants of exceptionally high electron density~\cite{zhang1999x, tikhonov2001phase}. Studies on hydrocarbon-based water-soluble surfactants merely addressed the influence of surfactant adsorption on interface roughness and tension~\cite{mcclain1994x, lee1996interfacial} but did not determine $\Gamma$.\\
In the present work, we demonstrate that $\Gamma$ can be extracted from XRR and GIXS curves when the analysis is aided by molecular dynamics (MD) simulations that correctly capture surfactant in-plane arrangements, conformations, and hydration. For water-insoluble lipids, such a procedure has already been proven successful~\cite{grava2023combining, suarez2025consistent}. Here, we focus on two technologically important yet chemically very different non-ionic surfactants without any elemental labels, the oligo-ethylene-glycol (OEG) surfactant \ce{C12EO6} and the glycosurfactant $\beta$-\ce{C12G2}. The latter has been shown to form adsorption layers with pronounced in-plane cohesion, which is beneficial for the stability of foams~\cite{stubenrauch2017hydrogen,  ranieri2018influence, kanduvc2021intersurfactant}. We find that XRR in combination with MD simulations yields saturation coverages consistent with NR when the contribution of capillary wave roughness on the interface is considered. Analysis of XRR and GIXS data leads to consistent results. Additional insights are gained by integrating the obtained adsorption isotherm with the \emph{equation~of~state} (EoS), $\gamma(\Gamma)$, which is accessible in MD but not in experiments.

\begin{figure}[H]
	\centering
	\includegraphics[width=0.43\textwidth]{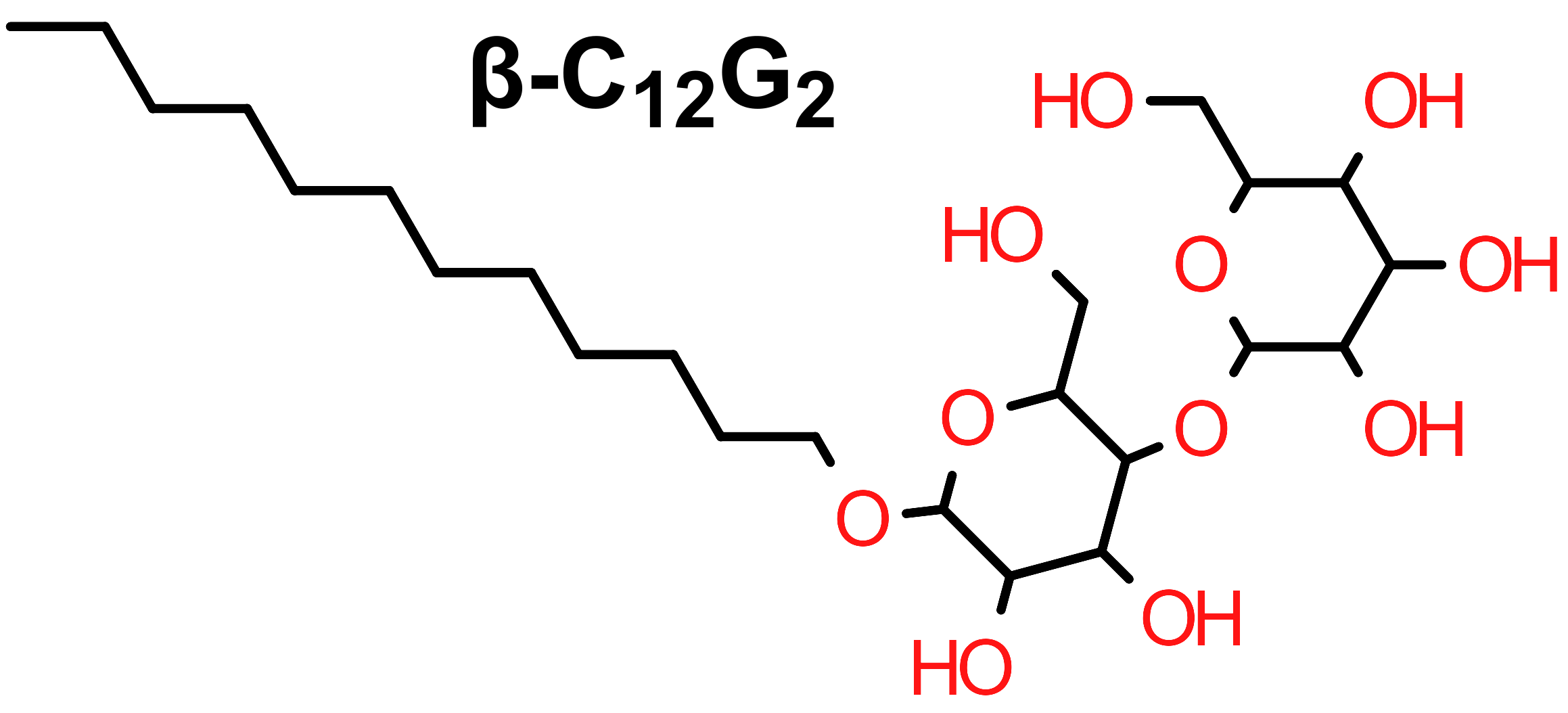}
    \includegraphics[width=0.7\linewidth]{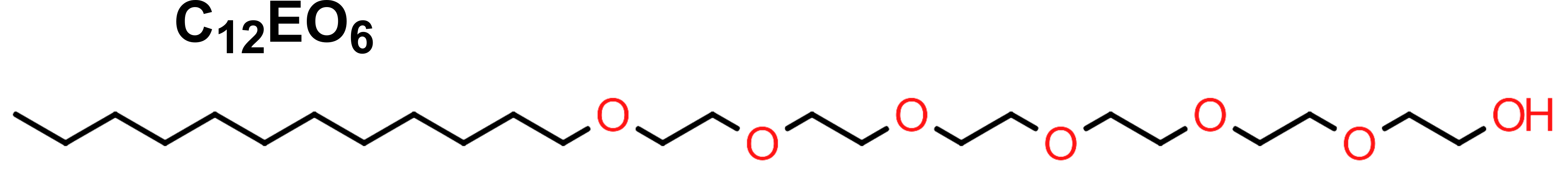}
	\caption{Chemical structures of the surfactants investigated.}
	\label{fig:structures}
\end{figure}

\section{Results and discussion}
The chemical structures of the two surfactants investigated are shown in Fig.~\ref{fig:structures}. Both comprise a saturated \ce{C12} alkyl chain and a non-ionic hydrophilic headgroup. The headgroup of hexa(oxyethylene)-lauryl-ether (\ce{C12EO6}) is a flexible, chain-like linear OEG with six repeat units with only one hydrogen-bond (HB) donor per surfactant. In contrast, n-dodecyl-$\beta$-D-maltoside ($\beta$-\ce{C12G2}) carries a comparatively rigid disaccharide headgroup with numerous HB donors and acceptors, which were demonstrated to form inter-surfactant HBs~\cite{stubenrauch2017hydrogen, kanduvc2021intersurfactant} that contribute to in-plane cohesion.  
 
\subsection{\ce{C12EO6}}
\begin{figure}[htb]
    \centering
    \includegraphics[width=.5\linewidth]{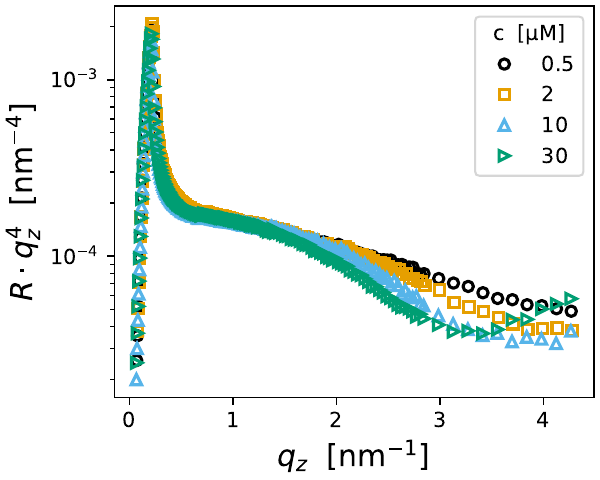}
    \caption{XRR curves $R(q_z)$ of the surfaces of aqueous \ce{C12EO6} solutions with bulk concentrations ranging from \qtyrange{0.5}{30}{\micro M}, all below the CMC of $\approx\qty{70}{\micro M}$\cite{patil2008binary}. For clarity, the curves are plotted as $R\cdot q_z^4$ on a logarithmic scale as a function of $q_z$.}
    \label{fig:xrr_measurements_c12eo6}
\end{figure}

Fig.~\ref{fig:xrr_measurements_c12eo6} shows X-ray reflectivity curves $R(q_z)$ of the interface between air and aqueous solutions of \ce{C12EO6} at systematically increasing bulk concentrations $c$, ranging from \qtyrange{0,5}{30}{\micro M}, all below the critical micelle concentration (CMC\, $\approx\qty{70}{\micro M}$, ref.~\citenum{patil2008binary}). In this regime, the surface coverage $\Gamma$ increases monotonically with increasing bulk concentration~\cite{rosen2012surfactants}. Surfactant adsorption typically leads to the formation of distinct layers in the electron density profile, thereby giving rise to reflectivity minima called Kiessig fringes~\cite{daillant2008x}, whose $q_z$ positions depend on the layer thickness. Indeed, the reflectivity curves in Fig.~\ref{fig:xrr_measurements_c12eo6} exhibit systematic changes with increasing concentration $c$, primarily concerning the appearance of a reflectivity minimum that gradually moves toward lower $q_z$. In other words, the reflectivity curves encode information about the layer thickness and, therefore, also about $\Gamma$. But even though it is possible to extract the approximate electron density profiles from the reflectivity data with slab-based models~\cite{kanduvc2021intersurfactant, schneck2024experimental, reed2025grazing}, these alone do not allow reconstructing $\Gamma$ in a reliable manner because electron density contributions from the surfactants and the displaced water molecules cannot simply be disentangled.

\begin{figure}[htb]
    \centering
    \includegraphics[height=10cm,
                    trim=0cm 5cm 0cm 3cm,
                    clip]{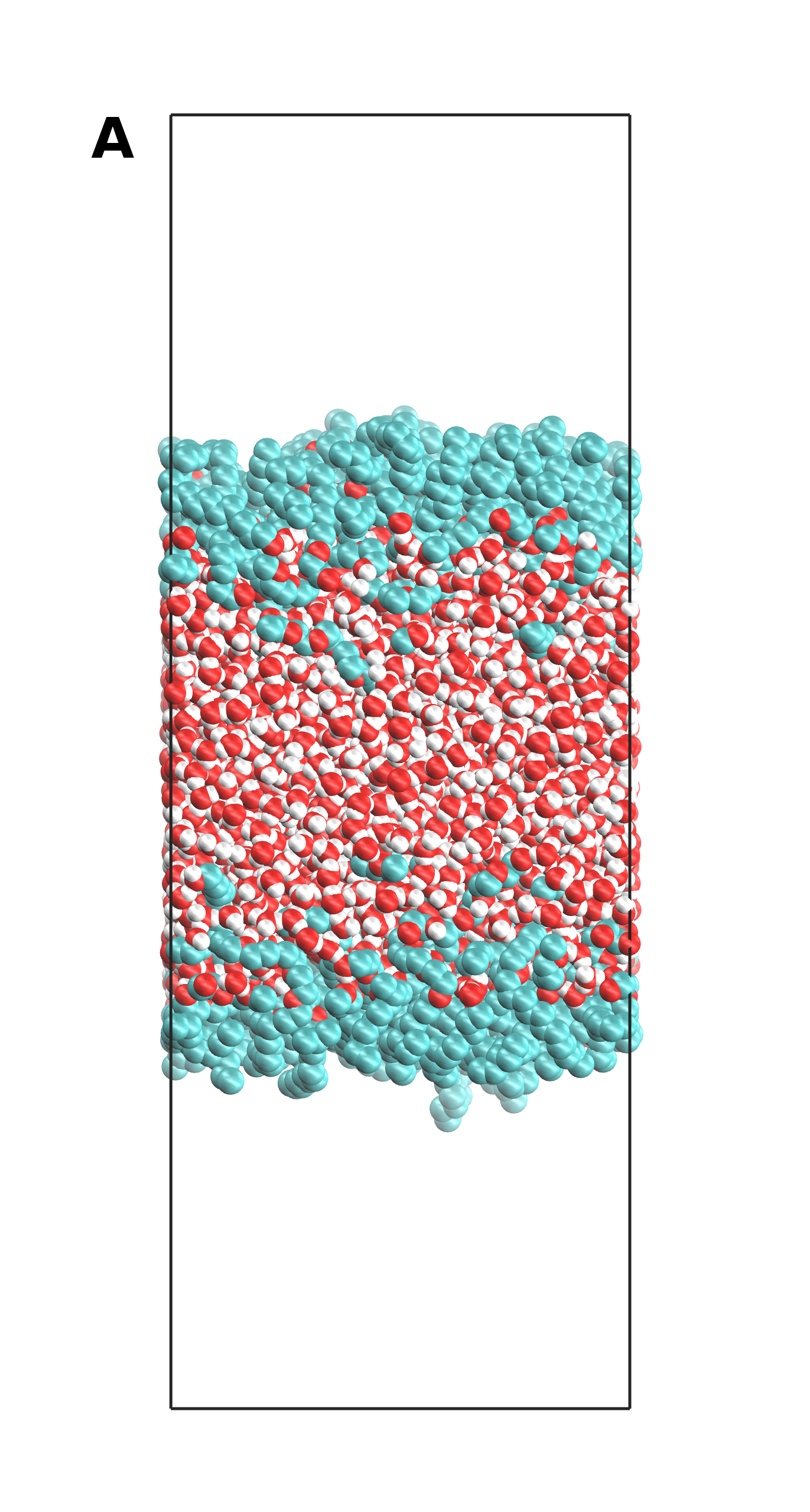}%
    \includegraphics[height=10cm]{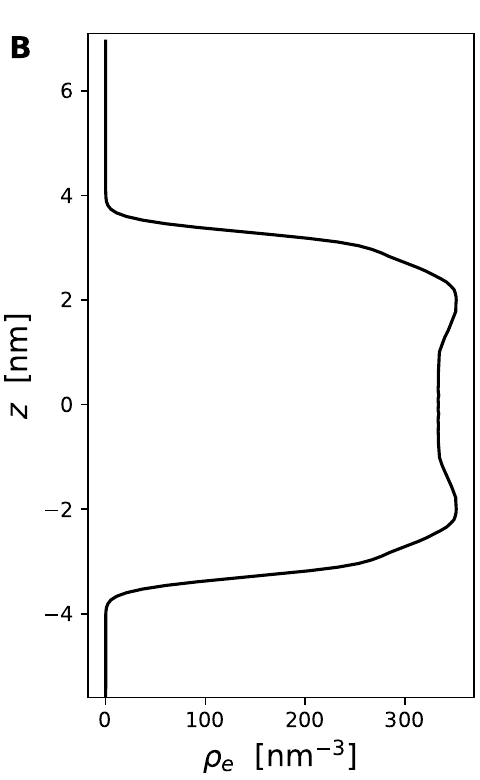}
    \caption{(A) Simulation snapshot of a water layer with \num{48} \ce{C12EO6} surfactants on each surface, corresponding to $\Gamma=\qty{2.0}{\per\nano\meter\squared}$. \ce{CH2} groups and \ce{CH3} groups (both represented as united atoms) are shown in light blue, oxygen atoms in red, and polar hydrogen atoms in white. The periodic simulation box is indicated with a rectangle. (B) Associated electron density profile $\rho_{\text{e}}(z)$ in the direction perpendicular to the interfaces.}
    \label{fig:simulation_snapshot}
\end{figure}

By contrast, MD simulations readily allow us to impose a desired value of $\Gamma$ by placing the corresponding number of surfactants at the air/water interface in the simulation box with a fixed area, as has previously been demonstrated for lipids~\cite{grava2023combining, suarez2025consistent}. $\Gamma$ then stays constant over the simulation time (here: \qty{300}{ns}, see Methods section) because exchange with the aqueous bulk occurs on much longer time scales~\cite{kanduc2023interface}. Fig.~\ref{fig:simulation_snapshot}A shows a snapshot from such a simulation (see the Methods section for technical details). The simulation box contains a water layer of about \qty{3.8}{\nano\meter} thickness whose two interfaces with the vacuum (representing air) are covered with a prespecified number of surfactants (here: 48 \ce{C12EO6} molecules per interface, corresponding to $\Gamma=\qty{2.0}{\per\nano\meter\squared}$), which are pinned there because of the hydrophobic effect. The molecules are oriented with their hydrophilic headgroups towards the water layer, as imposed by their amphiphilic character~\cite{Muller2021}.
Fig.~\ref{fig:simulation_snapshot}B shows the associated electron density profile $\rho_{\text{e}}(z)$ across the simulation box in the $z$ direction. The water layer was made to be thick enough so that the center features bulk properties, as evidenced from the featureless electron density plateau in the middle of the profile, with $\rho_{\text{e}}=\qty{333}{e^-\per\nano\meter\cubed}$. This is in good agreement with the well-known experimental value ($\rho_{\text{e}}=n_{\text{e}}^{\text{W}}\rho_{\text{M}}^{\text{W}}N_{\text{A}}/M_{\text{W}}=\qty{334}{e^-\per\nano\meter\cubed}$, where $n_{\text{e}}^{\text{W}}=10$ is the number of electrons per water molecule, $M_{\text{W}}=\qty{18}{\gram\per\mole}$ is the molecular mass, $\rho_{\text{M}}^{\text{W}}=\qty{1000}{\kilogram\per\meter\cubed}$ is the mass density of liquid water, and $N_{\text{A}}$ is Avogadro's constant). 

\begin{figure}[htb]
    \centering
    \includegraphics[width=0.4\linewidth]{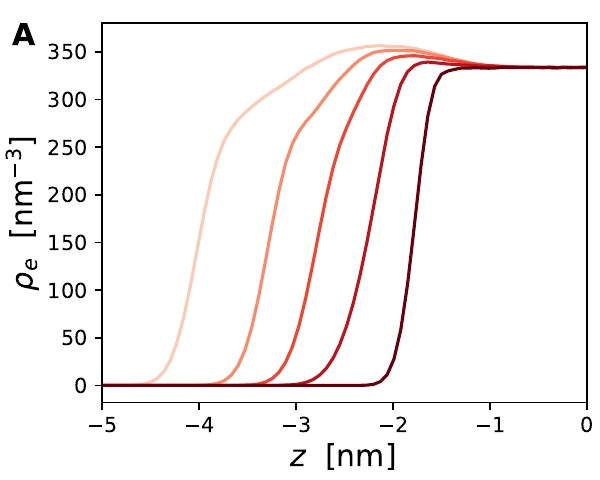}%
    \includegraphics[width=0.4\linewidth]{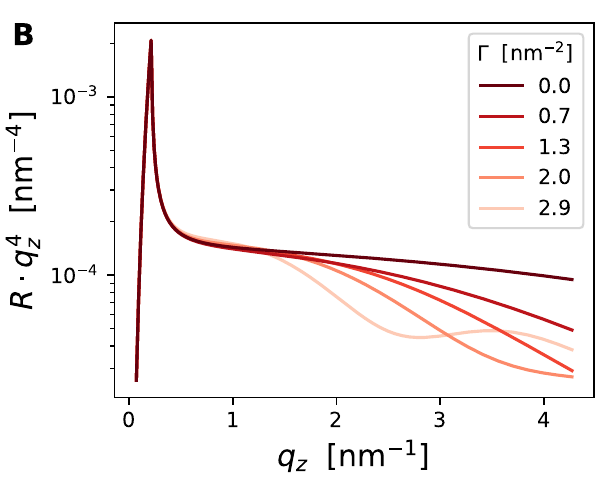}
    \caption{(A) One-sided electron density profiles from simulations with various surface coverages of \ce{C12EO6} surfactants (see legend in panel B). (B) Associated XRR curves. For clarity, the curves are plotted as $R\cdot q_z^4$ on a logarithmic scale as a function of $q_z$.}
    \label{fig:xrr_simulations}
\end{figure}

\begin{figure}[!htb]
    \centering
    \includegraphics[width=.45\linewidth]{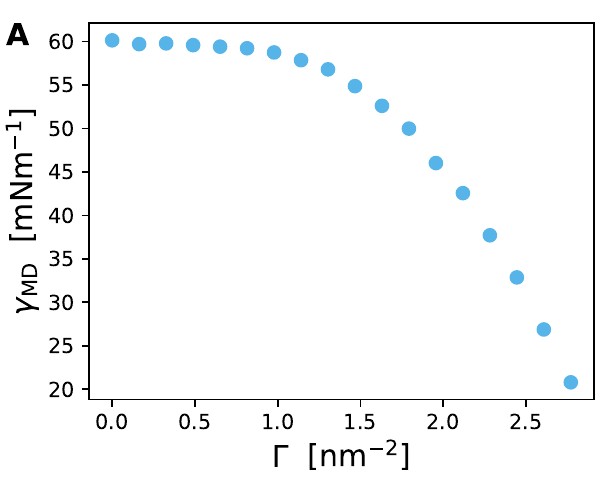}\hspace{7pt}
    \includegraphics[width=.45\linewidth]{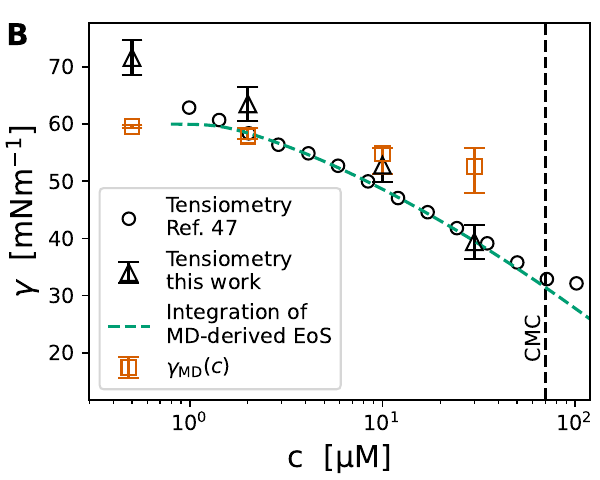}
    \caption{(A) Equation of state $\gamma_\mathrm{MD}(\Gamma)$ of \ce{C12EO6} as obtained in the simulations. Note that the water model used predicts a bare-water surface tension of $\gamma_0=\qty{60}{\milli\newton\per\meter}$. (B) Surface tension isotherm $\gamma_{\text{MD}}(c)$ reconstructed from the equation of state and from the simulation-assisted determination of the adsorption isotherm, together with independent tensiometric measurements from the present work and from the literature~\cite{angarska2007drainage}. Dashed line: $\gamma(c)$ as obtained through integration of the MD-derived equation of state (see Eq.~\ref{eq:gibbs_integrated}).}
    \label{fig:c12eo6_surface_tension}
\end{figure}

Fig.~\ref{fig:xrr_simulations}A shows the simulation-based one-sided electron density profiles that realistically represent single, surfactant-loaded interfaces between the two quasi-infinitely extended media, air and water. The profiles are computed from simulations with various choices of $\Gamma$, ranging from \qtyrange{0}{2.9}{\per\nano\meter\squared}, as imposed by the maximal surfactant number of \num{72} per interface in the simulation box. Higher coverages lead to unrealistically low surface tensions (see Fig.~\ref{fig:c12eo6_surface_tension}A) far below the experimental plateau surface tension above the CMC ($\approx\qty{30}{\milli\newton\per\meter}$)~\cite{angarska2007drainage}. In the absence of surfactants ($\Gamma=0$), the profile of the bare water surface exhibits a monotonic transition from $\rho_{\text{e}}=0$ to the bulk value mentioned before ($\rho_{\text{e}}=\qty{333}{e^{-}\per\nano\meter\cubed}$). As $\Gamma$ increases, the profiles change systematically, gradually developing a distinct headgroup-related density maximum between the hydrocarbon chain region and the water region. These systematic changes in the electron density profiles rationalize the systematic changes of the reflectivity curves in Fig.~\ref{fig:xrr_measurements_c12eo6} because the adsorption isotherm $\Gamma(c)$ increases monotonically, as mentioned before. Fig.~\ref{fig:xrr_simulations}B shows the theoretical XRR curves calculated from the $\Gamma$-dependent one-sided electron density profiles as described in the Methods section. The similarities with the experimental XRR curves in Fig.~\ref{fig:xrr_measurements_c12eo6} are striking. Most importantly, the trends when increasing the surface coverage in the theoretical XRR curves are consistent with those when increasing the bulk concentration in the experimental XRR curves. 

\begin{figure}[htb]
    \centering
    \includegraphics[width=0.45\linewidth]{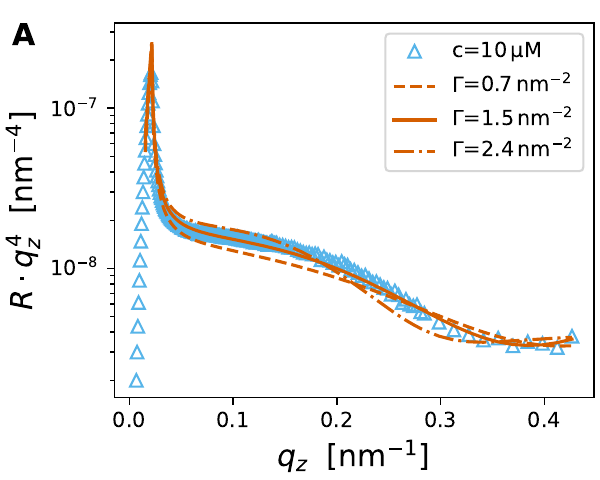}\quad%
    \includegraphics[width=0.45\linewidth]{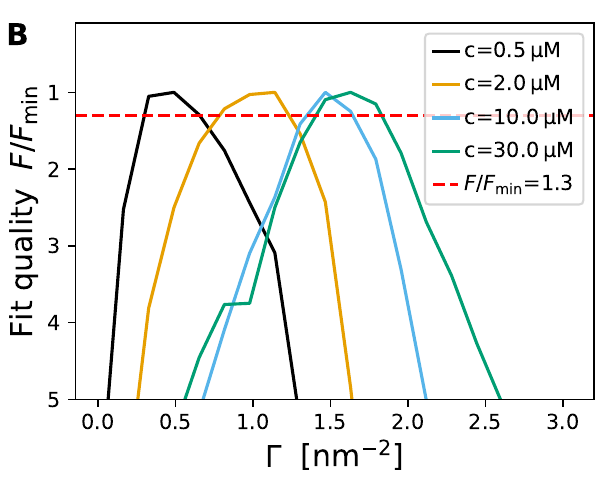}
    \caption{(A) Experimental XRR curve for \ce{C12EO6} at $c=\qty{10}{\micro M}$ (symbols) together with theoretical XRR curves (lines) corresponding to simulations with three different surface coverages. For clarity, the curves are plotted as $R\cdot q_z^4$ on a logarithmic scale as a function of $q_z$. Only $\Gamma=\qty{1.5}{\per\nano\meter\squared}$ leads to good agreement with the experimental reflectivity data. 
    (B) Normalized fit quality (mean squared difference, Eq.~\ref{eq:fitquality}) of the fits between the experimental and theoretical XRR curves for \ce{C12EO6}. The experimental XRR curves are recorded for four bulk concentrations (indicated in the legend) below the CMC of $\approx\qty{70}{\micro M}$\cite{patil2008binary}. The theoretical XRR curves are computed from simulations at different prespecified surface coverages $\Gamma$. The uncertainty threshold ($F\geq\num{1.3}\, F_\mathrm{min}$) is indicated with a dashed horizontal line.}
    \label{fig:c12eo6_bestfit_fitquality}
\end{figure}

In order to determine the surface coverages that match best the experimental XRR curves obtained at various bulk concentrations, we performed for all the different $\Gamma$ values a systematic comparison with the theoretical XRR curves of the form
\begin{equation}\label{eq:refl_simulation}
R(q_z)=sR_{\text{MD}}(q_z)\exp{(- q_z^2\sigma_{\Delta}^2)}+I_{\text{bg}},
\end{equation}
where $R_{\text{MD}}(q_z)$ is the reflectivity curve corresponding to the MD-derived electron density profile, $s$ is a scaling factor to account for imperfect normalization of the lab reflectometer, and $I_{\text{bg}}$ is a constant background. The exponential factor is a roughening correction explained below. The fit quality $F$ of each $\Gamma$ value was the mean squared deviation between the experimental and theoretical XRR curves (Eq.~\ref{eq:fitquality}). Fig.~\ref{fig:c12eo6_bestfit_fitquality}A shows the overlay of a representative experimental XRR curve (\ce{C12EO6}, $c=\qty{10}{\micro M}$) with theoretical XRR curves for three different $\Gamma$ values, which exhibit considerable deviations from the experimental data.

An important aspect to be considered in this comparison is the influence of the interface roughness, to which one major contribution is the thermally-induced capillary wave (CW) roughness. In general, CW roughness has different magnitudes in the experiments and in the simulations, as explained in the Methods section. In brief, the magnitude increases with decreasing surface tension and with the probed in-plane length scale~\cite{Braslau1985, mitrinovic1999x, shen2024reconstructing}, which is different in the simulations (the box size) and in the experiments. Prior to the comparison with the experimental data, we therefore apply to the theoretical reflectivity curves an exponential factor $\exp{(-q_z^2\sigma^2)}$. This is equivalent to convoluting the electron density profiles with a Gaussian function of width $\sigma_{\Delta}$ chosen such that it reproduces the CW roughness $\sigma_{\text{XRR}}$ expected to occur on the length scale probed by XRR, for a given value of the surface tension:

\begin{equation}
\sigma{_\Delta}=\sqrt{\sigma_{\text{XRR}}^2(\gamma_{\text{exp}})-\sigma_{\text{MD}}^2(\gamma_{\text{MD}})}.
\end{equation}

Here, $\sigma_{\text{MD}}(\gamma_{\text{MD}})$ is the CW roughness expected to occur on the simulation length scale for the surface tension $\gamma_{\text{MD}}$ observed in the simulation for a given surface coverage. Note that $\gamma_{\text{MD}}$ in general differs from the experimental value $\gamma_{\text{exp}}$, especially when the surface coverage is very different. The influence of this procedure on the theoretical XRR curves is exemplified in the Supporting Information (Fig.~S1).

\begin{figure}[htb]
    \centering
    \includegraphics[width=0.5\linewidth]{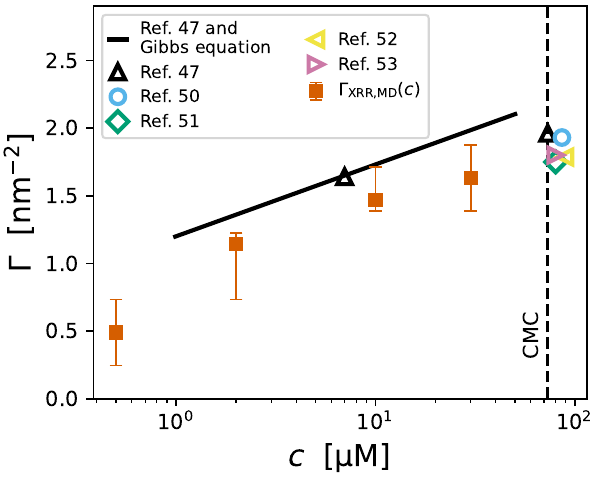}
    \caption{Adsorption isotherm $\Gamma_\mathrm{XRR,MD}(c)$ of \ce{C12EO6} as constructed from the simulation-assisted analysis of experimental XRR curves. The CMC is indicated with a dashed vertical line. Error bars correspond to the uncertainty regions in Fig.~\ref{fig:c12eo6_bestfit_fitquality}B. Solid line: adsorption isotherm reconstructed from tensiometric data by~\citealt{angarska2007drainage} with Eq.~\ref{eq:gibbs}. The plot also contains $\Gamma$ values reported in others studies\cite{angarska2007drainage, Eastoe1997, Zhmud2000, Yada2017, lu1993neutron}.}
    \label{fig:c12eo6_best_fits}
\end{figure}

Fig.~\ref{fig:c12eo6_bestfit_fitquality}B shows an overview of the normalized fit quality, $F/F_{\text{min}}$ (Eq.~\ref{eq:fitquality}) of the theoretical XRR curves to the experimental XRR curves obtained at all \ce{C12EO6} bulk concentrations. For each concentration, $F$ was calculated systematically for all $\Gamma$ values. The surface coverage that best agrees with the experimental data systematically increases with $c$. A plot of the adsorption isotherm $\Gamma_\mathrm{XRR,MD}(c)$ constructed by associating the best-matching surface coverage to each bulk concentration is shown in Fig.~\ref{fig:c12eo6_best_fits}. The error bars correspond to the uncertainty threshold indicated in Fig.~\ref{fig:c12eo6_bestfit_fitquality}B. $\Gamma$ is seen to increase monotonically with increasing bulk concentration, first approximately linearly with the logarithm of the concentration until it seems to turn to saturation as $c$ approaches the CMC. For the highest bulk concentration investigated ($c=\qty{30}{\micro M}<$ CMC), $\Gamma_\mathrm{XRR,MD}=\qty{1.6(2:2)}{\per\nano\meter\squared}$ is obtained, which is consistent with the value determined by NR~\cite{lu1993neutron} at the CMC, $\Gamma=\qty{1.8(2:2)}{\per\nano\meter\squared}$ and with $\Gamma\approx\qty{2.0}{\per\nano\meter\squared}$ estimated from tensiometry data~\cite{patil2008binary, angarska2007drainage}. The corresponding area per molecule, $A_{\text{mol}}=1/\Gamma_\mathrm{XRR,MD}\approx\qty{0.6}{\nano\meter\squared}$, is about three times the cross-sectional area of around $\qty{0.20}{\nano\meter\squared}$ required per stretched (all-trans) alkyl chain in a chain-crystalline arrangement~\cite{hermelink2008unsaturated}.
The significantly larger area requirement must be attributed to the fact that alkyl chains in adsorption layers of \ce{C12} surfactants are always melted and that the area requirement of the OEG headgroup is substantially larger because the loss of conformational entropy required for a stretched-out configuration cannot be compensated by direct hydrogen bonding~\cite{stubenrauch2017hydrogen}. Parameter-based thermodynamic models for the interpretation of tensiometric $\gamma(c)$ curves have predicted maximal coverages $\Gamma$ in the range of \qtyrange{1.6}{2.0}{\per\nano\meter\squared} for \ce{C12EO6}~\cite{Eastoe1997, Zhmud2000, Yada2017}. The solid line in Fig.~\ref{fig:c12eo6_best_fits} is the adsorption isotherm reconstructed from tensiometric data by~\citealt{angarska2007drainage} according to Eq.~\ref{eq:gibbs}, using a spline fit (see Supporting Information Fig.~S2). The concentration-dependent trend is consistent with the surface coverages determined in the present work, albeit with systematically slightly higher values.

Our interim result is that we have established the adsorption isotherm $\Gamma(c)$ of \ce{C12EO6}. The electron density profiles involved in this procedure can be considered robust with respect to the simulation force field because the relevant aspects concern molecular volumes and conformational flexibility, which are generally well captured by all force fields, as was pointed out recently for the case of lipids~\cite{suarez2025consistent}. As we will see further below, the quantitative results are, however, sensitive to the treatment of CW roughness. In the next step, we turn our attention to the equation of state $\gamma_\mathrm{MD}(\Gamma)$, shown in Fig.~\ref{fig:c12eo6_surface_tension}A. It is obtained from MD simulations by analyzing the pressure tensor~\cite{Muller2021}, as described in the Methods section. The tension $\gamma_\mathrm{MD}$ is seen to decay monotonically with increasing $\Gamma$, first slowly and then fast, when steric repulsion becomes dominant~\cite{kovalchuk2023surfactant}. With $\gamma_0=\qty{60}{\milli\newton\per\meter}$, the tension of the bare water surface (at $\Gamma=0$) produced by the employed water model is significantly lower than the experimental value ($\gamma_0=\qty{72}{\milli\newton\per\meter}$). It should also be noted that the equation of state is more sensitive to the force field than the electron density profiles from which we reconstructed the adsorption isotherm because the former has a subtle dependence on polar and non-polar interaction parameters. We would therefore like to assess how well it reproduces the adsorption layer's equation of state. Unfortunately, we do not have a purely experimental curve $\gamma(\Gamma)$ at hand because of the limitations mentioned in the Introduction. However, we can construct from $\Gamma_\mathrm{XRR,MD}(c)$ the surface tension isotherm $\gamma_\mathrm{MD}(c):=\gamma_\mathrm{MD}(\Gamma_\mathrm{XRR,MD}(c))$, which can be compared to available experimental $\gamma(c)$ data and constitutes an indirect manifestation of the equation of state~\cite{kanduc2023interface}.
Fig.~\ref{fig:c12eo6_surface_tension}B compares $\gamma_{\text{MD}}(c)$ with experimental data measured tensiometrically by us (see Methods section) and by~\citealt{angarska2007drainage}. In the Supporting Information (Fig.~S4), the data are also shown in terms of the surface pressure $\Pi=\gamma_0-\gamma$. The agreement between the predictions of the combined XRR and simulation approach for $\gamma(c)$ and the direct tensiometric measurements of $\gamma(c)$ is not very good, particularly regarding the slope of the curve. We therefore assess the extent to which this deviation is attributable to the equation of state generated by the simulation force field. For this purpose, we reconstruct $\gamma(c)$ from the simulations through a different route, by directly integrating the EoS according to Eq.~\ref{eq:gibbs} (see also Methods section):
\begin{equation}\label{eq:gibbs_integrated}
\ln{c}=\ln{c_0}-\frac{1}{k_{\text{B}}T}\int_{\gamma(c_0)}^{\gamma(c)}\frac{1}{\Gamma(\gamma')}\text{d}\gamma',    
\end{equation}
where $c_0$ is an unspecified reference concentration. The numerical integration of the MD-derived EoS works well as long as $\gamma(c_0)$ is very close to $\gamma_0$, which is safely fulfilled for the lowest non-zero surface coverages realized in the simulations. The resulting curve $\gamma(c)$, obtained by numerically inverting Eq.~\ref{eq:gibbs_integrated}, is plotted in Fig.~\ref{fig:c12eo6_surface_tension}B as a dashed line. Note that the concentration scale is determined only up to the unknown value of $c_0$, which can, however, be obtained through a fit to the experimental data (here: $c_0=\qty{0.79}{\micro M}$). At first glance, the shape of the MD-derived EoS seems consistent with the experimental tension isotherm, indicating that the simulation force field captures the surfactants' in-plane interactions acceptably well. We note, however, that there are ambiguities related to the undetermined concentration scale in combination with the deviating value of $\gamma_0$ in the simulations. An additional factor for the unsatisfactory prediction of $\gamma(c)$ by the combined XRR and simulation approach may be a slight systematic underestimation of the surface coverage by our approach. As shown in the Supporting Information (Fig.~S5), $\Gamma_{\text{XRR,MD}}$ is systematically shifted to lower values when larger values of $\sigma_{\Delta}$ (Eq.~\ref{eq:refl_simulation}) are imposed. Slightly overestimated roughness corrections would thus lead to substantially underestimated reductions of the surface tensions, which would be consistent with the mismatch observed in Fig.~\ref{fig:c12eo6_surface_tension}B. Possible systematic uncertainties in the calculation of $\sigma_{\Delta}$ include uncertainties in the effectively probed in-plane length scale in XRR and MD, the different boundary conditions, as well as inconsistencies between the surface bending rigidities in experiments and simulations or center-of-mass drift correction and symmetrization in MD. In addition, surface roughness contributions not captured by the CW model, such as kinetic arrest or systematic distortion, may have different manifestations in simulations and experiments. These aspects will therefore have to be systematically investigated.

\subsection{$\beta$-\ce{C12G2}}
\begin{figure}[htb]
    \centering
    \includegraphics[width=0.45\linewidth]{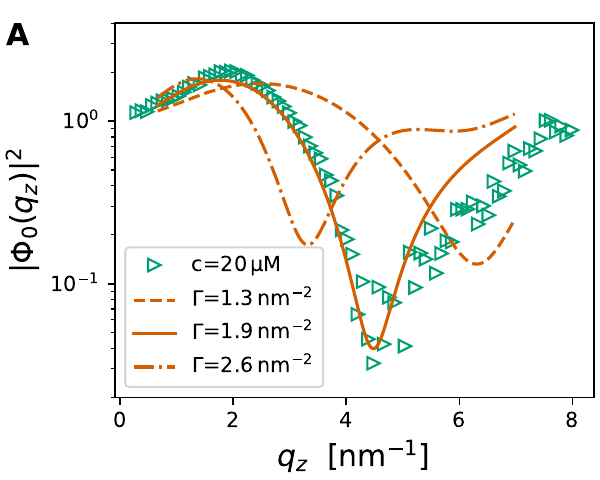}\quad%
    \includegraphics[width=0.45\linewidth]{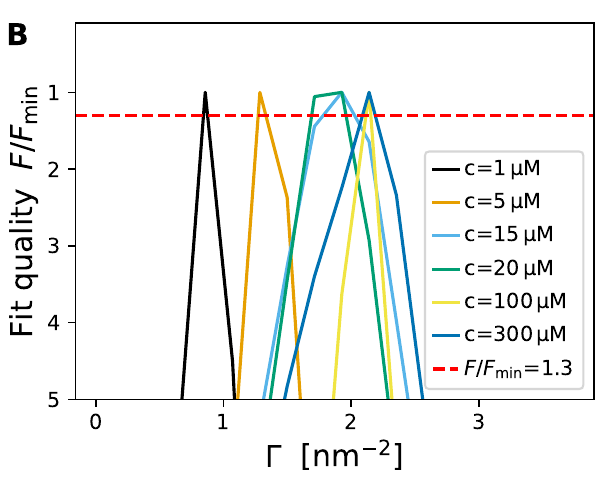}
    \caption{(A) Experimental $|\Phi_0(q_z)|^2$ curve for $\beta$-\ce{C12G2} at $c=\qty{20}{\micro M}$ (symbols) together with theoretical curves (lines) corresponding to simulations with three different surface coverages. For clarity, the curves are plotted on a logarithmic scale. Only $\Gamma=\qty{1.9}{\per\nano\meter\squared}$ leads to good agreement with the experimental data.
    (B) Normalized fit quality (mean squared difference, Eq.~\ref{eq:fitquality_pseudorefl}) of the fits between the experimental and theoretical XRR curves for $\beta$-\ce{C12G2}. The experimental XRR curves are recorded for six bulk concentrations (indicated in the legend) below the CMC of $\approx\qty{150}{\micro M}$\cite{patil2008binary}. The theoretical XRR curves are computed from simulations at different prespecified surface coverages $\Gamma$. The uncertainty threshold ($F\geq\num{1.3}\, F_\mathrm{min}$) is indicated with a dashed horizontal line.}
    \label{fig:c12g2_bestfit_fitquality}
\end{figure}

For the second surfactant investigated, $\beta$-\ce{C12G2} (with CMC~=~$\qty{150}{\micro M}$)~\cite{patil2008binary}, we use data from synchrotron-based GIXS, which offers a wider $q_z$ range. Here, we deal with the GIXS-derived squared modulus of the intrinsic structure factor $|\Phi_0(q_z)|^2$, which corresponds to the electron density profile of an idealized layer structure without broadening due to thermally-induced CW fluctuations~\cite{shen2024reconstructing} (see Methods section and Fig.~S6). Fig.~\ref{fig:c12g2_bestfit_fitquality}A shows $|\Phi_0(q_z)|^2$ for one bulk concentration, together with fits for three $\Gamma$ values realized in the simulations. In these fits, the roughness correction was implemented by modeling the experimental data as
\begin{equation}\label{eq:phi_simulation}
    |\Phi_0(q_z)|^2=\frac{R_\mathrm{MD}(q_z)}{R_F(q_z)}\frac{1}{\exp{\left(-q_z^2\sigma_{\text{MD}}^2(\gamma_{\text{MD}})\right)}} + I_\text{bg},
\end{equation}
where $R_{\text{MD}}(q_z)$ is the reflectivity curve corresponding to the MD-derived electron density profile, $R_{\text{F}}(q_z)$ is the reflectivity of an idealized (step-like) air/water interface. $I_{\text{bg}}$ is a constant background, and division by the exponential factor containing $\sigma_{\text{MD}}^2(\gamma_{\text{MD}})$ compensates broadening due to the predicted CW roughness occurring in the simulations (Fig.~S7). In Fig.~\ref{fig:c12g2_bestfit_fitquality}B, the fit quality, as defined in the Methods section, is shown for the six bulk concentrations.

\begin{figure}[htb]
    \centering
    \includegraphics[width=.45\linewidth]{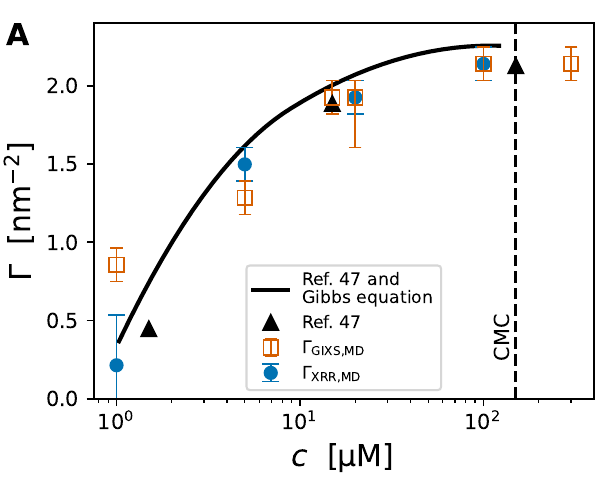}
    \includegraphics[width=.45\linewidth]{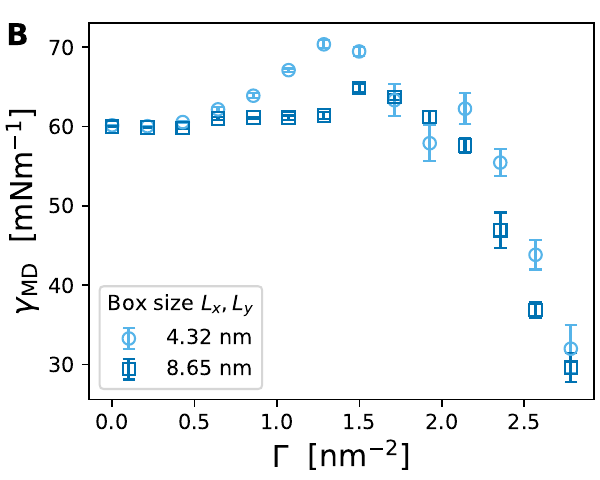}
    \caption{(A) Adsorption isotherm $\Gamma(c)$ of $\beta$-\ce{C12G2} as constructed from the simulation-assisted analysis of experimental GIXS and XRR curves. The CMC is indicated with a dashed vertical line. Error bars correspond to the uncertainty regions in Fig.~\ref{fig:c12g2_bestfit_fitquality} (GIXS) and in Fig.~S9 (XRR). Solid line: adsorption isotherm reconstructed via Eq.~\ref{eq:gibbs} from tensiometric data by~\citealt{angarska2007drainage} (see Fig.~S3). The plot also includes $\Gamma$ values reported in this study. (B) Equation of state $\gamma_{\text{MD}}(\Gamma)$ of $\beta$-\ce{C12G2} as obtained in simulations with different box sizes.}
    \label{fig:c12g2_best_fits}
\end{figure}

The resulting adsorption isotherm $\Gamma(c)$ is shown in Fig.~\ref{fig:c12g2_best_fits}A (open squares). It increases monotonically with $c$ and exhibits a tendency of saturation at around $\Gamma\approx\qty{2}{\per\nano\meter\squared}$ as $c$ approaches the CMC, in line with the expectation. The plot also contains surface coverage data deduced from complementary XRR measurements (Fig.~S8 and S9) of $\beta$-\ce{C12G2} in the same way as described above for \ce{C12EO6}. Reassuringly, the coverages deduced from XRR are in good agreement with those deduced from GIXS. The adsorption isotherm reconstructed from tensiometric data by~\citealt{angarska2007drainage} according to Eq.~\ref{eq:gibbs} (solid line), yields overall consistent values.

The simulation-derived equation of state of $\beta$-\ce{C12G2} is shown in Fig.~\ref{fig:c12g2_best_fits}B for two different simulation box sizes. Unlike the monotonically decaying curve for \ce{C12EO6}, $\beta$-\ce{C12G2} features a non-monotonic trend with a maximum of $\gamma$ at intermediate coverages of $\Gamma\approx\qty{1.5}{\per\nano\meter\squared}$. This ``hump'', with values exceeding the surface tension of pure water, corresponds to negative surface pressures. Such values indicate metastable states of either a continuous monolayer under lateral tension or a condensed monolayer with a pore. The latter leads to a greater increase in tension for smaller pores (as shown in the Supporting Information), which rationalizes the more pronounced hump in the smaller simulation box. Both metastable scenarios in the simulations correspond to a two-phase coexistence of gas and condensed phase in the macroscopic limit. This means that, for comparison with experiments, the humps in Fig.~\ref{fig:c12g2_best_fits}B would have to be replaced with a constant surface tension slightly below $\gamma_0$, followed by a decrease at $\Gamma\approx \qty{2}{\per\nano\meter\squared}$ once the surface becomes fully covered by a condensed monolayer. The fact that this hump is observed for $\beta$-\ce{C12G2} but not for \ce{C12EO6} points to substantially stronger cohesion between $\beta$-\ce{C12G2} surfactants. It is most likely mediated by hydrogen bonding between their disaccharide headgroups, as was suggested earlier~\cite{stubenrauch2017hydrogen, ranieri2018influence, kanduvc2021intersurfactant}. Whether this cohesion is strong enough to drive a genuine gas/condensed transition in reality remains unclear. The lateral cohesion is probably over-expressed in the simulation model, which is a typical shortcoming of current force fields~\cite{tempra2022accurate}. Such a transition would imply a discontinuous increase in the surface coverage $\Gamma(c)$ at the coexistence bulk concentration. According to Eq.~\ref{eq:gibbs}, it would also cause a corresponding kink in the surface tension isotherm $\gamma(c)$ as observed for some other surfactants~\cite{aratono1984phase} but not as decisively for $\beta$-\ce{C12G2}~\cite{patil2008binary}. Nevertheless, we consider it plausible that $\beta$-\ce{C12G2} exhibits some significant clustering tendency, albeit likely less pronounced than with the force field used for the simulations. For the sake of completeness we remark that, due to finite sampling and statistics, we cannot exclude that also the equation of state of \ce{C12EO6} (Fig.~\ref{fig:c12eo6_surface_tension}A) exhibits minimal rudiments of such non-monotonicity.

\section{Conclusions}
To quantify the concentration-dependent adsorption of surfactants at air/water interfaces, we combined X-ray reflectivity with molecular dynamics simulations, which provide an accurate description of the coverage-dependent interfacial electron density profile. In this way, we were able to determine $\Gamma(c)$ for two common non-ionic surfactants, \ce{C12EO6} and $\beta$-\ce{C12G2}. We found consistency when analyzing XRR and GIXS data, and our results are in satisfactory agreement with previous NR results on the saturation surface coverage. The adsorption isotherms obtained in this way can provide a valuable basis for interpreting tensiometric data using parameter-based thermodynamic models and for examining force fields for molecular dynamics simulations. For both surfactants, we observe deviations between the experimental tension isotherms and those predicted by the combined XRR and MD approach. For $\beta$-\ce{C12G2}, this must be attributed to the strong lateral cohesion of the surfactants, which is likely overexpressed by the force field and leads to finite-size artifacts in MD. For \ce{C12EO6}, we hypothesize that the deviations result from an imperfect representation of the equation of state in the simulations, combined with a slight underestimation of surface coverage by our approach, with the main uncertainty arising from the subtleties of roughness matching between experiments and simulations. Future work should therefore focus on systematic studies to provide a basis for a more rigorous quantitative treatment of the capillary-wave roughnesses effectively probed by XRR and MD simulations, in addition to other roughness contributions. Finally, improved simulation force fields of the surfactants and water would be beneficial for a reliable and precise measurement of $\Gamma(c)$, also with laboratory X-ray sources.

\section{Materials and methods}

\subsection{Chemicals and sample preparation}
\ce{C12EO6} and $\beta$-\ce{C12G2} were purchased from Merck (Darmstadt, Germany) and from Glycon (Luckenwalde, Germany), respectively, and used as received. \ce{H2O} was ultrapure Milli-Q water (specific resistivity: \qty{18}{\mega\ohm\centi\meter}) from a purification system (Elga, Purelab classic). Aqueous solutions of the surfactants at the desired concentrations were prepared by adding the respective surfactant amounts to \ce{H2O} in glass bottles previously cleaned by extensive rinsing with ethanol and \ce{H2O}. The solutions were then stirred and left to rest for at least \num{30}~min before being poured into the troughs used for XRR or GIXS measurements.  

\subsection{Surface tension measurements}
The surface tension was measured in a Teflon trough by Riegler \& Kirstein (Potsdam, Germany) with a Wilhelmy paper plate. Because no \emph{in situ} calibration measurement with pure water is possible, the uncertainty in $\gamma$ was estimated as $\qty{\pm 2}{\milli\newton\per\meter}$. As seen in Fig.~\ref{fig:c12eo6_surface_tension}B, the measured surface tensions are consistent with earlier reports of the equilibrium surface tension, indicating that equilibration was reached when the measurements were performed.

\subsection{XRR}
XRR measurements were performed at room temperature, \qty{20 \pm 1}{\celsius}, using a D8 Advance reflectometer (Bruker AXS, Karlsruhe, Germany) with a horizontal sample geometry and a vertical goniometer, allowing the liquid surface to be studied without disturbance during the measurements. A Teflon trough (Riegler \& Kirstein GmbH, Potsdam, Germany), with the dimensions of \qty{11.5}{\centi\meter} parallel and \qty{32.5}{\centi\meter} perpendicular to the beam propagation, was enclosed in a box with Kapton windows through which the incident and reflected X-ray beams pass. XRR curves were measured in the $\theta-2\theta$ geometry, where $\theta$ is the incident angle. A conventional X-ray tube with a copper anode (Cu K$_{\alpha}$, wavelength $\lambda = \qty{0.154}{\nano\meter}$) was used to generate an X-ray beam with a line focus. The beam was monochromatized with a Göbel mirror (W/Si multilayer) and collimated through two narrow horizontal slits of \qty{0.1}{\milli\meter} width with a switchable Cu attenuator in between. Soller slits ($\Delta\theta_x = \qty{25}{\milli\radian}$) were placed after the last horizontal slit and directly in front of the detector.
Detection resolution is defined by a $0.1\,\mathrm{mm} \times 15\,\mathrm{mm}$ (vertical $\times$ horizontal) slit at $220\,\mathrm{mm}$, and the intensity is integrated over this slit by the line detector. The intensity was recorded with a Våntec-1 line detector (Bruker AXS, Germany). Intensities were corrected using known attenuation factors and, where necessary, for the variable footprint. Finally, the angular XRR curves were transformed to XRR curves as a function of the perpendicular scattering vector component, $q_z=4\pi \sin(\theta/\lambda)$. Laboratory-based XRR has extremely low intensity and therefore comes without risk of beam damage or heating. Moreover, the fast exchange of the surfactants between the interface and the bulk ensures that beam damage does not accumulate over time.

\subsection{GIXS}
Grazing incidence X-ray scattering (GIXS) data as a function of the vertical scattering angle $\beta$ were measured at the beamline P08~\cite{seeck2012high} of PETRA III using the Langmuir trough GID setup at $\qty{15}{\kilo\electronvolt}$ at an incident angle $\alpha= \qty{0.07}{\degree}$ \cite{shen2022grazing} at a horizontal scattering angle $2\theta=\qty{0.3}{\degree}$ off the place of incidence. This angle was defined by two post-sample vertical slits~\cite{dai2011comparative}, and corresponded to $q_{xy}=\qty{0.04}{\per\angstrom}$ at $\beta=0$. The angular resolutions $\Delta2\theta$ and $\Delta\beta$ were both $\qty{0.08}{\degree}$ (FWHM). The Langmuir trough was located in a hermetically sealed container with Kapton windows, and the temperature was kept at $\qty{20}{\celsius}$ by a thermostat. The container was constantly flushed with a stream of humidified helium (He) to prevent air scattering and the generation of reactive oxygen species. A ground glass plate was placed approximately $\qtyrange{0.3}{1}{\milli\meter}$ beneath the illuminated area of the monolayer in order to reduce mechanically excited surface waves. GIXS has negligible beam damage thanks to the spreading of flux over a large footprint, short measurement time (140~s), the use of 15keV, and the use of helium atmosphere.

Subtracting the minimal intensity from the GIXS data as background yields the surface diffuse scattering $R^{*}\left(q_{z},q_{xy}\right)$. The squared modulus $|\Phi_{0}\left(q_{z}\right)|^2$ of the intrinsic structure factor was then calculated from $R^{*}\left(q_{z},q_{xy}\right)$ as
\begin{equation}
|\Phi_{0}\left(q_{z}\right)|^2=\frac{R^{*}\left(q_{z},q_{xy}\right)}{\Psi\left(q_{z},q_{xy}\right)|_{\gamma, T}}\cdot\frac{\left(2q_{z}\right)^4}{q_\text{c}^4|t_{\alpha}|^2|t_{\beta}|^2},
\end{equation}
where $q_\text{c}$, $t_{\alpha}$, and $t_{\beta}$ are the critical angle of the water surface and the transmission coefficients of the incident and the scattered waves, respectively. The roughness factor $\Psi\left(q_{z},q_{xy}\right)|_{\gamma, T}$ was calculated as~\cite{shen2024reconstructing}
\begin{equation}
\Psi\left(q_{z},q_{xy}\right)|_{\gamma, T}=\frac{q_{z}^4}{16\pi^2\sin{\alpha}}\cdot q_{\text{max}}^{-\eta}\cdot\frac{k_{\text{B}}T}{\gamma}\int_{2\theta-\frac{\Delta2\theta}{2}}^{2\theta+\frac{\Delta2\theta}{2}}\int_{\beta-\frac{\Delta\beta}{2}}^{\beta+\frac{\Delta\beta}{2}}\frac{\cos{\beta}\text{d}\beta\text{d}2\theta}{q_{xy}^{2-\eta}}
\end{equation}
where $\gamma$ is the surface tension, $k_{\text{B}}$ the Boltzmann constant, $T$ the temperature, $\eta=k_{\text{B}}Tq_{z}^2/(2\pi\gamma)$, and $q_{\text{max}}=\pi/a_{\text{m}}$ is the reciprocal space cutoff for the average molecular length scale ($a_{\text{m}}\approx\qty{0.5}{\nano\metre}$). The intrinsic structure factor corresponds to the electron density profile of an idealized layered structure, without broadening from thermally induced capillary-wave fluctuations.

\subsection{Capillary wave roughness prediction}
The CW roughness $\sigma_{\text{CW}}$ of a liquid surface is directly related to the in-plane length $L_\text{p}$ of the surface being probed \cite{Pershan2000,shen2025extending}:
\begin{equation}
    \sigma_{\text{CW}}^2 \approx \frac{k_{\text{B}} T}{2 \pi \gamma} \ln \frac{q_{\text{max}}}{\delta q_{xy}} - \frac{k_{\text{B}} T}{2 \pi \gamma} K_0 \left( \frac{q_\kappa}{q_{\text{max}}} \right),
\end{equation}
where $\delta q_{xy} = L_\text{p}^{-1}$ is the in-plane resolution (HWHM) in reciprocal space. The large-$q$ cutoff $q_{\text{max}} = \pi / a_{\text{m}}$ is related to the molecular size $a_{\text{m}}$ (assumed here as $\qty{0.5}{\nano\metre}$), and the cutoff $q_\kappa = \sqrt{\gamma / \kappa}$ depends on the ratio between the tension $\gamma$ and the surface bending modulus $\kappa$. $K_0(x)$ is the modified Bessel function of the $2^{\text{nd}}$ kind of the $0^{\text{th}}$ order. In MD simulations, $L_\text{p}$ coincides with the box dimension. The effect of the bending rigidity was neglected due to the low modulus ($\kappa\lesssim3k_{\text{B}}T$, see Supporting Information, Fig.~S10), so that we are left with:

\begin{equation}
    \sigma_{\text{CW}}^2 \approx \frac{k_{\text{B}} T}{2 \pi \gamma} \ln \frac{q_{\text{max}}}{\delta q_{xy}}.
\end{equation}

\subsection{Molecular dynamics simulations}
Parts of the following section are replicated from previous publications\cite{kanduvc2021intersurfactant, kanduc2023interface}. For \ce{C12EO6}, we used the GROMOS-compatible 2016H66 force field \cite{Horta2016, Senac2017} and for $\beta$-\ce{C12G2}, we used the GROMOS 53a6 force field \cite{Lopez2013, VanEerden2015, Oostenbrink2004}. Both surfactant force fields adopt a united-atom treatment of the aliphatic groups. Water was described with the SPC/E model~\cite{Berendsen1987}. The simulation setup consisted of a \qtyrange{3}{4}{\nano\meter} thick water slab replicated in the $x$ and $y$ directions via periodic boundary conditions, placed in a simulation box of dimensions around \qtyproduct[product-units=single]{4.3x4.3x12}{\nano\meter} for $\beta$-\ce{C12G2} and around \qtyproduct[product-units=single]{5x5x14}{\nano\meter} for \ce{C12EO6}. The water slab was loaded with the same number of surfactants on its two surfaces. Variable surfactant coverages were achieved by symmetrically varying the number of surfactants on both surfaces while leaving the box dimensions fixed. The simulation topology and run files are available in the supporting information\cite{tudatalib}. The simulations were performed with the GROMACS simulation package \cite{VanDerSpoel2005, Abraham2015}, version 2021.5. Electrostatics was treated using the particle--mesh-Ewald method \cite{Darden1993, Essmann1995} with a \qty{1.4}{\nano\meter} real-space cutoff. The Lennard-Jones interactions were cut off at \qty{1.4}{\nano\meter}. The simulations were performed with an integration time step of \qty{2}{fs} in the constant-volume ($NVT$) ensemble. Temperature was maintained at \qty{293.15}{K} using the velocity-rescale thermostat \cite{Bussi2007} with a time constant of \qty{0.1}{ps}. Each simulation was \qty{300}{ns} long. However, following initial investigations of common measures of equilibration, the first \qty{100}{ns} were discarded from the analysis. The surface tension was calculated from the diagonal elements $p_{xx}$, $p_{yy}$, and $p_{zz}$ of the pressure tensor, as $\gamma=L_z[2p_{zz}-(p_{xx}+p_{yy})]/(2N_{\text{int}})$, where $L_z$ is the simulation box size in $z$ direction and $N_{\text{int}}=2$ is the number of interfaces in the simulation box \cite{Muller2021}.

\subsection{Electron density profiles from the MD simulations}
Electron density profiles $\rho_{\text{e}}(z)$ were obtained from molecular dynamics simulations with the "gmx density" tool of GROMACS~\cite{Abraham2015}. For this procedure, center-of-mass subtraction and symmetrization were used to enable averaging over multiple time steps and to improve statistics, respectively. The $z$-axis was divided into slabs of equal thickness (\qty{0.07}{\nano\meter} for \ce{C12EO6} and \qty{0.06}{\nano\meter} for $\beta$-\ce{C12G2}). 

\subsection{Calculation of theoretical X-ray reflectivity curves}
With the discretized set of electron density slabs at hand, the partial waves reflected at each slab/slab interface were computed according to the Fresnel equations and summed in a phase-correct manner using the Parratt recursion formalism \cite{Parratt1954} to determine the reflected intensity under idealized conditions, $R_{\text{MD}}(q_z)$. For comparison with the experimental curves, an offset $\Delta q_z$ was applied to account for minor angular misalignments that affect the precise apparent position of the critical angle of total reflection. 

\subsection{Fitting procedure and fit quality assessment}
The starting point is the combination of an experimental and a theoretical X-ray reflectivity curve, where the latter depends on the choice of $\Gamma$ in the simulation. To optimize the match between the two curves, the parameters $s$ and $I_{\text{bg}}$ in Eqs.~\ref{eq:refl_simulation} and \ref{eq:phi_simulation} had to be adjusted. Since the best-matching value $\Delta q_z$ is characteristic of each experimental curve and independent of $\Gamma$, it was constrained prior to the actual fit. The remaining free parameters for each choice of $\Gamma$ are then $s$ and $I_{\text{bg}}$ (or only $I_{\text{bg}}$). They were optimized in a classical least-squares minimization procedure. The fit quality function $F$ was chosen such that the fit is sufficiently sensitive to the high-$q_z$-features that encode the coverage.

For XRR, the fit quality function minimized was 
\begin{equation}\label{eq:fitquality}
F=\dfrac{1}{N}\sum_{i=1}^N w_i\left[\log\left(R_{i}^{\text{th}} q_{z,i}^4\right) - \log\left(R_{i}^{\text{ex}} q_{z,i}^4\right)\right]^2,
\end{equation}
where $R^{\text{th}}$ and $R^{\text{ex}}$ are the theoretical and experimental reflectivities, respectively, $N$ is the number of data points (i.e., the number of $q_z$ values at which the reflectivity was measured),
and $w_i$ is a weighting factor to compensate for non-uniform sampling along the $q_z$-axis. It was assigned proportionally to the local spacing between adjacent data points.

For fits of $|\Phi_0(q_z)|^2$, the quality function minimized was
\begin{equation}\label{eq:fitquality_pseudorefl}
F=\dfrac{1}{N}\sum_{i=1}^N \left[\log\left(|\Phi_0^{\text{th}}(q_{z,i})|^2\right) - \log\left(|\Phi_0^{\text{ex}}(q_{z,i})|^2 \right)\right]^2,
\end{equation}
where $|\Phi_0^{\text{th}}(q_{z,i})|^2$ and $|\Phi_0^{\text{ex}}(q_{z,i})|^2$ are the theoretical and experimental data, respectively. The fitting range was limited to the consensus range of validity ($q_z>3 \cdot q_z^{\text{c}}$, where $q_z^{\text{c}}$ is the $q_z$-value at the critical angle of total reflection).

The uncertainty margin for the parameters $\Gamma$ and $\sigma$ was defined as the point at which the deviation $F$ increases by a factor of \num{1.3} with respect to the optimal parameter set. This experience-based definition is somewhat arbitrary but, in our opinion, reflects the true uncertainty much better than estimates based solely on the statistical errors~\cite{bevington1993data}, which are much smaller but neglect systematic uncertainties, as pointed out earlier~\cite{rodriguez2017neutron}.  

\section{Supporting information}
Influence of the roughening parameter $\sigma_\Delta$ on reflectivity;
Splines of tensiometric data;
\ce{C12EO6} surface pressure isotherm;
Influence of $\sigma_\Delta$ on $\Gamma_{\text{XRR,MD}}$ for \ce{C12EO6};
Experimental and theoretical $|\Phi_0|^2$ data for $\beta$-\ce{C12G2};
XRR data and fit quality for $\beta$-\ce{C12G2};
Influence of the system size on the surface tension in a simulation containing a condensed monolayer with a pore;
Bending rigidity for $\beta$-\ce{C12G2}.

\section{Conflicts of interest}
There are no conflicts of interest to declare.

\section{CRediT author statement}
Conceptualization: MK, BL, ES; Formal Analysis: KD, CS; Investigation: KD, JR, DM; Methodology: KD, CS, MK, ES; Project administration: ES, BL; Resources: BL, ES; Supervision: ES, BL; Validation: KD, JR, CS, MK, ES; Visualization: KD; Writing – original draft: KD, JR, ES; Writing – review \& editing: KD, JR, CS, MK, BL, ES.

\begin{acknowledgement}
We acknowledge DESY (Hamburg, Germany), a member of the Helmholtz Association HGF, for the provision of experimental facilities. Parts of this research were carried out at PETRA III, and we would like to thank Chen Shen and Rene Kirchhof for assistance with P08 and Milena Lippmann for assistance in the chemistry lab. Beamtime was allocated for proposal I-11015005. M.K. acknowledges financial support from the Slovenian Research and Innovation Agency ARIS (contracts P1-0055, J1-4382, and J1-70038).	We thank Olaf Soltwedel and Hayden Robertson for support with XRR experiments and Reinhard Miller for insightful comments.
\end{acknowledgement}

\bibliography{references}

\clearpage
\vspace*{\fill}
\begin{center}
    \includegraphics[width=0.9\linewidth]{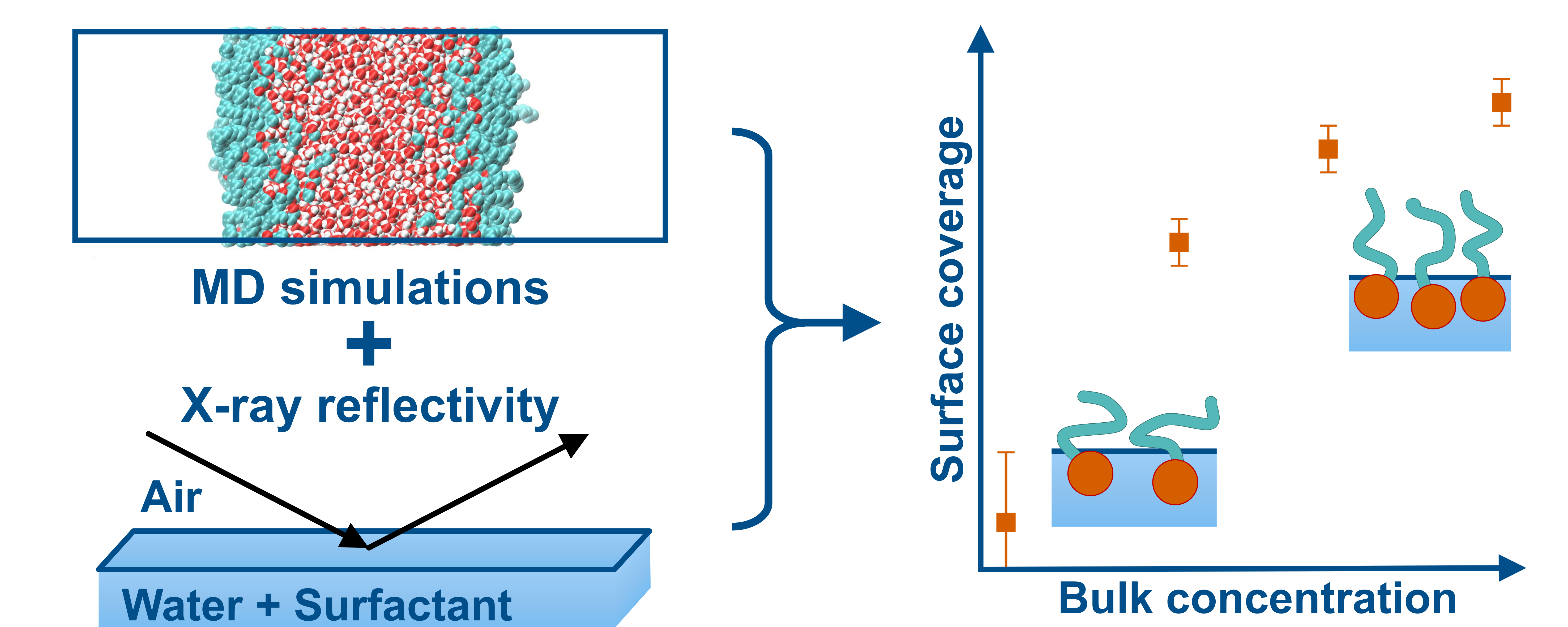}
\end{center}
\vspace*{\fill}
\clearpage

\end{document}


\maketitle

\newpage
\section{Influence of the roughening parameter $\sigma_\Delta$ on reflectivity}

\begin{figure}[htb]
    \centering
    \includegraphics[width=0.4\linewidth]{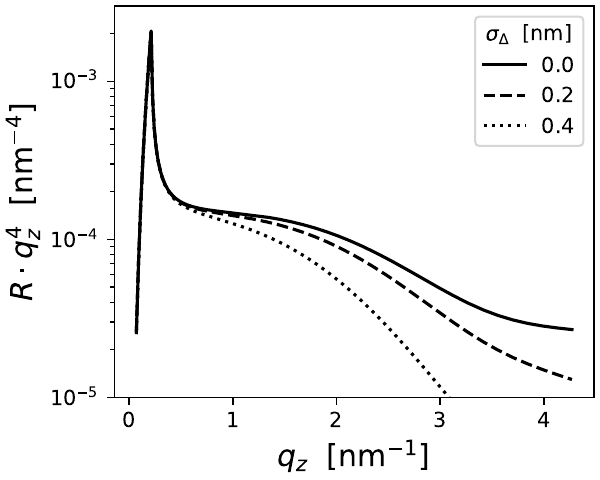}
    \caption{Influence of the roughening parameter $\sigma_\Delta$ on the theoretical reflectivity exemplified for \ce{C12EO6} at $\Gamma=\qty{2.0}{\per\nano\meter\squared}$. Original reflectivity curve (solid line), $\sigma_\Delta=\qty{0.2}{\nano\meter}$ (dashed line), and $\sigma_\Delta=\qty{0.4}{\nano\meter}$ (dotted line). Zero background intensity is assumed.}
    \label{fig:xrr_simulations_convoluted}
\end{figure}

\section{Splines of tensiometric data}

\begin{figure}[H]
    \centering
    \includegraphics[width=0.5\linewidth]{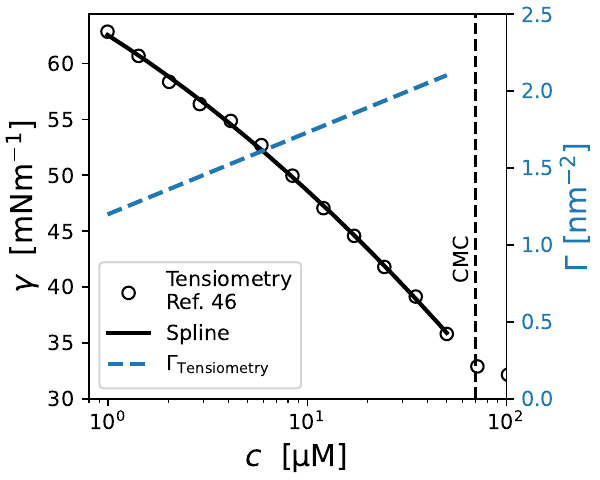}
    \caption{Tensiometry data for \ce{C12EO6} with a spline of second (quadratic) order. The dashed line is the surface excess $\Gamma$ determined from the spline with the Gibbs equation (see main text).}
    \label{fig:c12eo6_spline}
\end{figure}

\begin{figure}[H]
    \centering
    \includegraphics[width=0.5\linewidth]{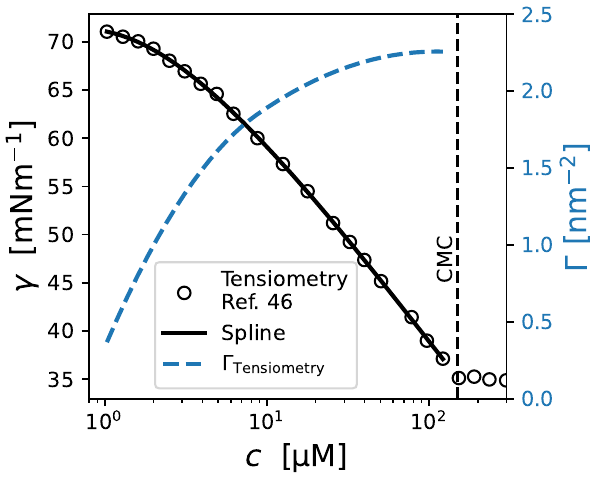}
    \caption{Tensiometry data for $\beta$-\ce{C12G2} with a spline of third (cubic) order. The dashed line is the surface excess $\Gamma$ determined from the spline with the Gibbs equation (see main text).}
    \label{fig:c12g2_spline}
\end{figure}

\section{\ce{C12EO6} surface pressure isotherm}
\begin{figure}[htb]
    \centering
    \includegraphics[width=.45\linewidth]{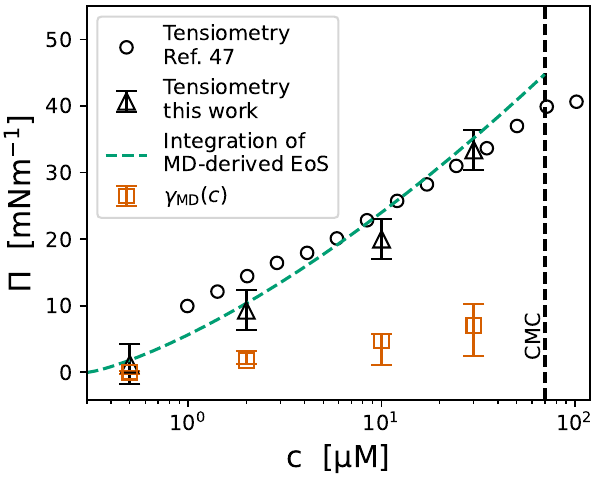}
    \caption{Surface pressure isotherm $\Pi(c)$ of \ce{C12EO6} reconstructed from the equation of state and from the simulation-assisted determination of the adsorption isotherm, together with independent tensiometric measurements from the present work and from the literature (see main text). Dashed line: $\Pi(c)$ as obtained through integration of the MD-derived equation of state with $c_0=0.17$\,µM (see main text).}
    \label{fig:c12eo6_absolute_surface_tension}
\end{figure}

\section{Influence of $\sigma_\Delta$ on $\Gamma_{\text{XRR,MD}}$ for \ce{C12EO6}}
\begin{figure}[!htb]
    \centering
    \includegraphics[width=0.7\linewidth]{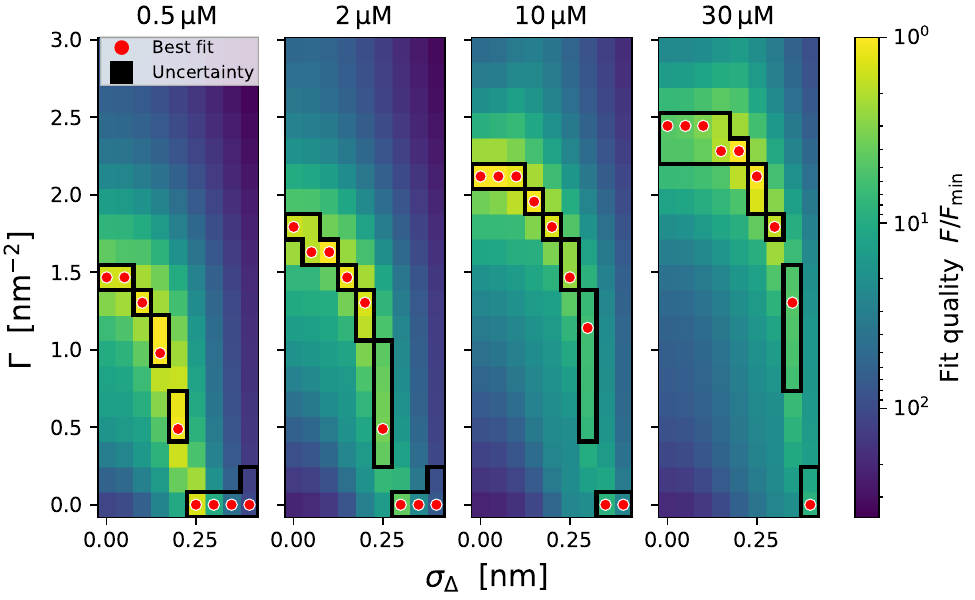}
    \caption{Heat map of the normalized fit quality $F/F_\mathrm{min}$ (mean squared difference, see main text) of the fits between the experimental and theoretical XRR curves for \ce{C12EO6}. The experimental XRR curves are recorded for four bulk concentrations (indicated on the top) below the CMC of $\approx\qty{70}{\micro M}$. The theoretical XRR curves are computed from simulations at different prespecified surface coverages $\Gamma$ (vertical axis). The roughening parameter $\sigma_\Delta$ (horizontal axis) is applied to the MD results. The best-matching parameter combinations for each visualized $\sigma_\Delta$ are marked in red, while the corresponding uncertainty regions ($F\leq\num{1.3}\, F_\mathrm{min}$) are surrounded by black lines.}
    \label{fig:c12eo6_heatmap}
\end{figure}

The predicted values of $\sigma_\Delta$ for the roughness matching of XRR and MD are shown in Table~\ref{tab:roughness_values}. In Fig.~\ref{fig:c12eo6_heatmap}, the fit quality for different $\sigma_\Delta$ is visualized as heatmaps. It can be seen that, for $\sigma_\Delta$ values lower than predicted, the best-fitting surface coverage $\Gamma$ increases.

\begin{table}[!ht]
    \centering
    \caption{Capillary wave roughness $\sigma_\mathrm{CW}$ for the four bulk concentrations $c$ of the experiments $\sigma_\mathrm{CW,XRR}$ and corresponding MD simulations $\sigma_\mathrm{CW,MD}$ with the best fits between the experimental and theoretical XRR curves for \ce{C12EO6}. For each bulk concentration, the necessary additional $\sigma_\Delta$ is also shown. See main text for details.}
    \label{tab:roughness_values}
    \begin{tabular}{c c c c}
        {$c$ [\unit{\micro M}]} & {$\sigma_{\text{CW,XRR}}$ [\unit{\nano\meter}]} & {$\sigma_{\text{CW,MD}}$ [\unit{\nano\meter}]} & {$\sigma_{\Delta}$ [\unit{\nano\meter}]} \\
        \hline
        0.5    & 0.28 & 0.19 & 0.21 \\
        2.0    & 0.30 & 0.20 & 0.23 \\
        10.0   & 0.33 & 0.20 & 0.26 \\
        30.0   & 0.38 & 0.20 & 0.32 \\
        \hline
    \end{tabular}
\end{table}

\FloatBarrier
\newpage

\section{Experimental and theoretical $|\Phi_0|^2$ data for $\beta$-\ce{C12G2}}
\begin{figure}
    \centering
    \includegraphics[width=0.5\linewidth]{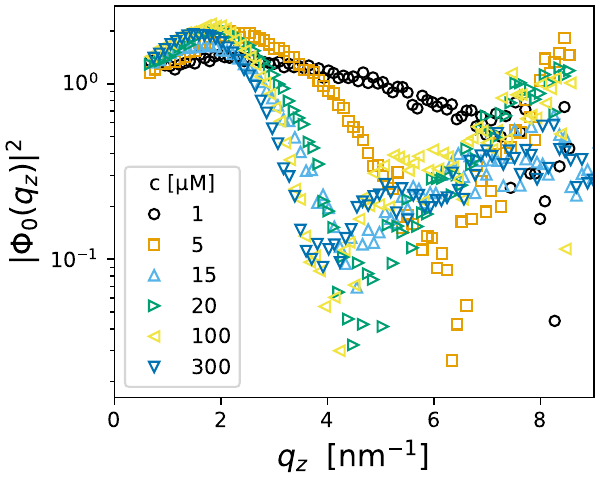}%
    \caption{Experimental $|\Phi_0|^2$ data for $\beta$-\ce{C12G2} from GIXS measurements at different bulk concentrations $c$ below the CMC of $\qty{150}{\micro M}$ (see main text).}
    \label{fig:c12g2_phi_exp}
\end{figure}

\begin{figure}
    \centering
    \includegraphics[width=0.45\linewidth]{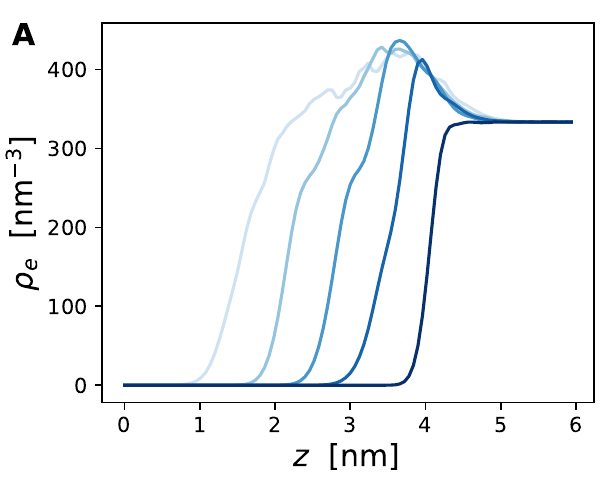}\quad%
    \includegraphics[width=0.45\linewidth]{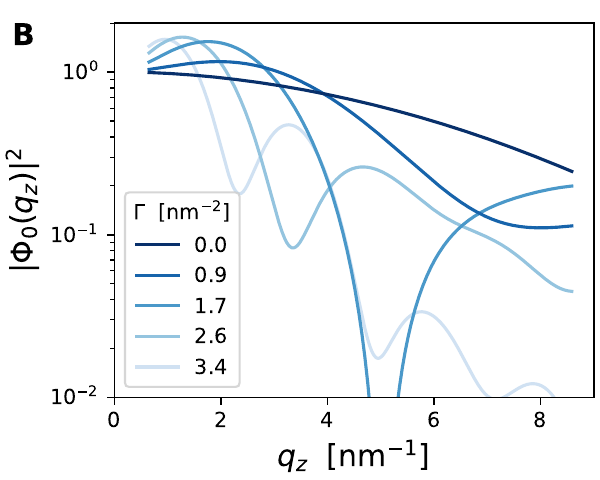}
    \caption{(A) Electron density profiles from MD simulations of $\beta$-\ce{C12G2} at different surface coverages (see legend in panel B). (B) Corresponding theoretical $|\Phi_0|^2$ data with $\sigma_\Delta=0$.}
    \label{fig:c12g2_edens_phi}
\end{figure}

Fig.~\ref{fig:c12g2_phi_exp} shows the experimental $|\Phi_0|^2$ data obtained with solutions of $\beta$-\ce{C12G2} at the concentrations indicated in the figure legend. The electron density profiles obtained in simulations for $0\leq\Gamma\leq\qty{3.4}{\per\nano\meter\squared}$ are shown in Fig.~\ref{fig:c12g2_edens_phi}A. The associated theoretical $|\Phi_0|^2$ curves are shown in Fig.~\ref{fig:c12g2_edens_phi}B.

\FloatBarrier
\newpage

\clearpage

\section{XRR data and fit quality for $\beta$-\ce{C12G2}}
\begin{figure}
    \centering
    \includegraphics[width=0.45\linewidth]{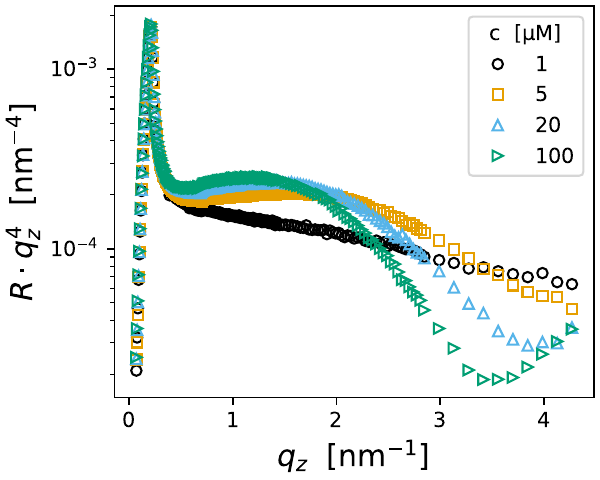}
    \caption{XRR curves $R(q_z)$ of the surfaces of aqueous $\beta$-\ce{C12G2} solutions with bulk concentrations ranging from $1$ to $\qty{100}{\micro M}$, all below the CMC of $\approx\qty{150}{\micro M}$. For clarity, the curves are plotted as $R\cdot q_z^4$ on a logarithmic scale as a function of $q_z$.}
    \label{fig:xrr_c12g2}
\end{figure}

\begin{figure}
    \centering
    \includegraphics[width=0.45\linewidth]{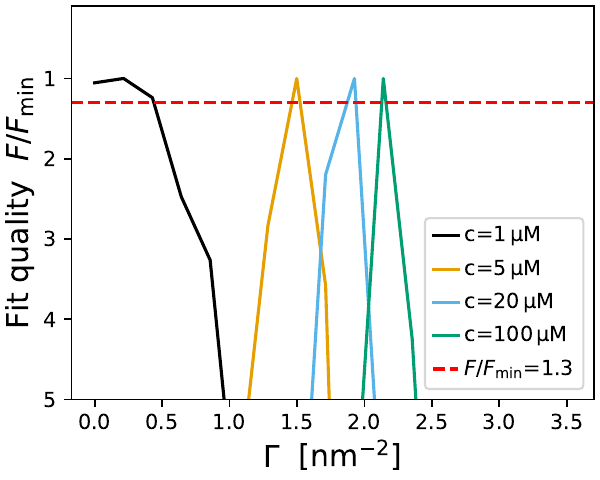}
    \caption{Normalized fit quality (mean squared difference, see main text) of the fits between the experimental and theoretical XRR curves for $\beta$-\ce{C12G2}. The experimental XRR curves are recorded for three bulk concentrations (indicated in the legend) below the CMC of $\approx\qty{150}{\micro M}$. The theoretical XRR curves are computed from simulations at different prespecified surface coverages $\Gamma$. The uncertainty threshold ($F\geq\num{1.3}\, F_\mathrm{min}$) is indicated with a dashed horizontal line.}
    \label{fig:c12g2_xrr_fitquality}
\end{figure}

Fig.~\ref{fig:xrr_c12g2} shows the XRR data for $\beta$-\ce{C12G2} for four bulk concentrations.
The fit qualities for XRR on $\beta$-\ce{C12G2} at these bulk concentrations are shown in Fig.~\ref{fig:c12g2_xrr_fitquality}.

\FloatBarrier

\newpage

\section{Influence of the system size on the surface tension in a simulation containing a condensed monolayer with a pore}
For a condensed surfactant monolayer with a circular pore, the surface tension increment $\Delta\gamma$ generated through the pore's line tension $\tau$ is inversely proportional to the pore radius $R$. This follows from the identity
\begin{equation}
\Delta\gamma = \frac{\mathrm{d}G}{\mathrm{d}A} = \frac{\mathrm{d}G}{\mathrm{d}R}\cdot\frac{\mathrm{d}R}{\mathrm{d}A} = \frac{\mathrm{d}G/\mathrm{d}R}{\mathrm{d}A/\mathrm{d}R} = \frac{2\pi\tau}{2\pi R} = \frac{\tau}{R},
\end{equation}
where $L = 2\pi R$ and $A = \pi R^{2}$ are the circumference and the area of the pore, respectively, and $G=\tau L$ is the excess free energy of the line. For two systems of different sizes but the same surface coverage $\Gamma$, the larger system will have a larger pore. Thus, the pore-related tension increment $\Delta\gamma$ will be smaller in the larger system.

\clearpage
\section{Bending rigidity for $\beta$-\ce{C12G2}}
\begin{figure}
    \centering
    \includegraphics[width=0.5\linewidth]{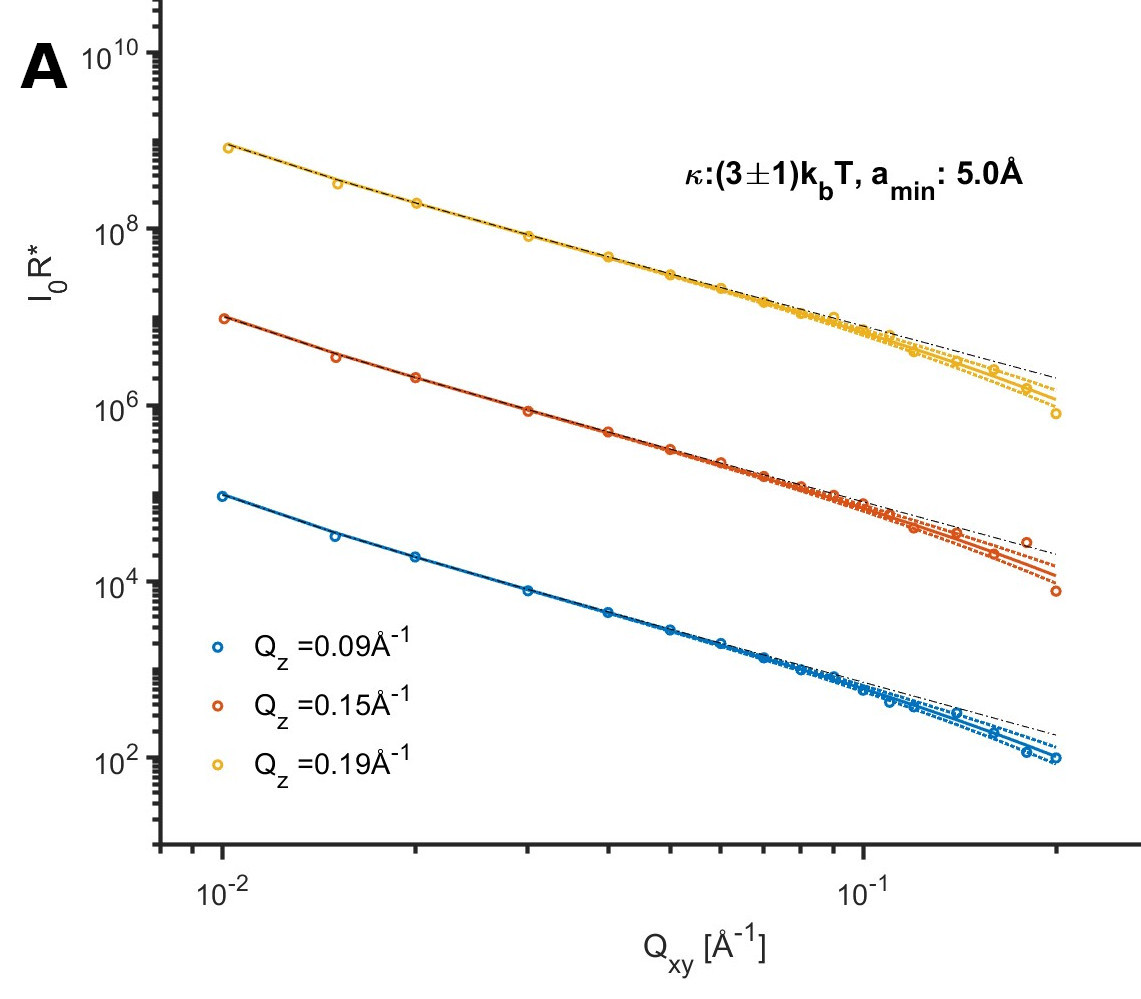}%
    \includegraphics[width=0.5\linewidth]{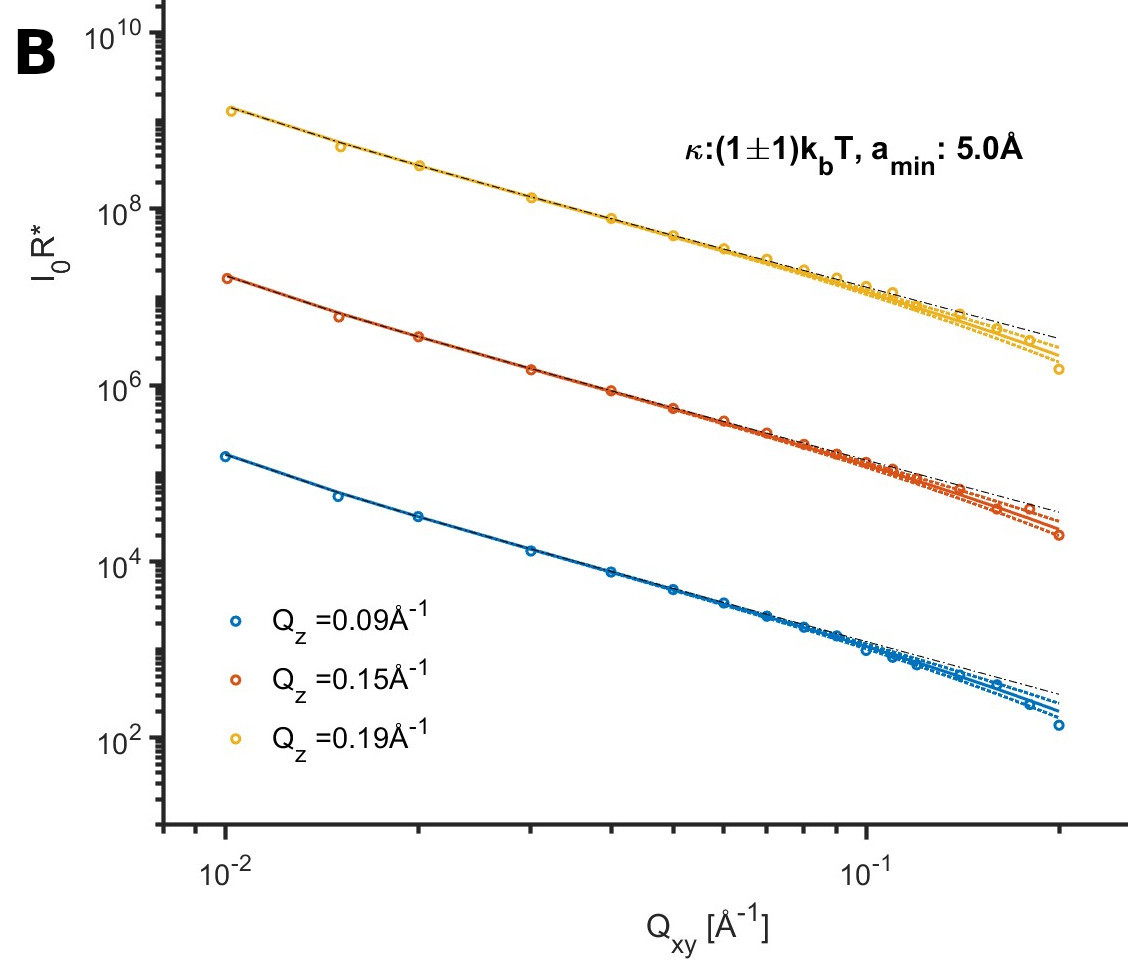}%
    \caption{Bending rigidity estimates for $\beta$-\ce{C12G2} for (A) $\qty{15}{\micro M}$ and (B) $\qty{300}{\micro M}$}
    \label{fig:c12g2_bending}
\end{figure}

Fig.~\ref{fig:c12g2_bending} shows estimates for the bending rigidity of $\beta$-\ce{C12G2} for two bulk concentrations. These were made on the basis of the $q_{xy}$-dependent diffuse reflectivity with the method described in Ref. 60.

